\documentclass[%
 reprint,
 amsmath,amssymb,
 aps,nofootinbib,   
]{revtex4-1}

\usepackage{amsmath}
\usepackage{amssymb}

\usepackage{bm}
\PassOptionsToPackage{linktocpage}{hyperref}
\usepackage[hyperindex,breaklinks]{hyperref}
\usepackage{enumitem}
\usepackage{slashed}

\newcommand{\diff}{\mathrm{d}}
\usepackage{xcolor}
\usepackage{float}

\renewcommand*\vec[1]{\bm{#1}}

\newcommand{\komma}{\, ,}
\newcommand{\punkt}{\, .}
\newcommand{\im}{\mathrm{i}}
\newcommand{\abs}[1]{\left| #1 \right|} 

\newcommand{\logEdeltam}{\log\left(\frac{E\delta}{m}\right)}
\newcommand{\e}[1]{\text{e}^{#1}}
\newcommand{\kommentar}[1]{}

\newcommand{\thetas}{{\theta'}}
\newcommand{\phis}{{\phi'}}
\newcommand{\thetaq}{{\theta_q}}
\newcommand{\phiq}{{\phi_q}}
\newcommand{\omegas}{{\omega'}}
\newcommand{\omegaq}{{\omega_q}}
\newcommand{\varepsilons}{{\varepsilon'}}
\newcommand{\varepsilonq}{{\varepsilon_q}}
\newcommand{\lambdas}{{\lambda'}}
\newcommand{\lambdaq}{{\lambda_q}}
\newcommand{\us}{{u'}}

\usepackage{array}
\usepackage{mathtools}

\usepackage{etoolbox}

\begin{document} 

\title{Masses and electric charges: gauge anomalies and anomalous thresholds}

\author{  C\'esar G\'omez and Raoul Letschka} 
\affiliation{Instituto de F\'{i}sica Te\'orica UAM-CSIC, Universidad Aut\'onoma de Madrid, Cantoblanco, 28049 Madrid, Spain}

\begin{abstract}
We work out in the forward limit and up to order $e^6$ in perturbation theory the collinear divergences. In this kinematical regime we discover new collinear divergences that we argue can be only cancelled using quantum interference with processes contributing to the gauge anomaly. This rules out the possibility of a quantum consistent and anomaly free theory with massless charges and long range interactions. We use the anomalous threshold singularities to derive a gravitational lower bound on the mass of the lightest charged fermion.

\end{abstract}
\maketitle

\section{Introduction}
\noindent
For theories with long range gauge forces as QED the IR completion problem goes around the quantum consistency of a quantum field theory with massless charged particles in the physical spectrum. 
This is an old problem that has been considered from different angles along the years (see \cite{Case,Gribov,Frohlich,Morchio} for an incomplete list). As a matter of fact in Nature we don't have any example of massless charged particles. In the Standard Model this is the case both for spin $1/2$ as well as for the spin $1$ charged vector bosons. In the particular case of charged leptons the potential inconsistency of a massless limit should imply severe constraints on the consistency of vanishing Yukawa couplings. 

Technically the infrared (IR) origin of the problem is easy to identify. In the case of massless charged particles radiative corrections due to loops of virtual photons lead to two types of infrared problem. One can be solved, in principle, using the standard Bloch-Nordiesk-recipe \cite{YFS}  that leads to infrared finite inclusive cross sections at each order in perturbation theory depending on an energy resolution cutoff. In this case the infrared finite cross section is defined taking into account soft radiation. In the massless case we have in addition collinear divergences that contribute logarithmically to Weinberg's  $B$ factor \cite{YFS,Weinberg}.\footnote{For more recent discussions on IR divergences see for instance \cite{Strominger,us} and references therein.} The standard recipe used to cancel these divergences requires to include in the definition of the inclusive cross section not only soft emission but also collinear hard emission and absorption (i.e.~photons with energy bigger than the energy resolution scale) and to set an angular resolution scale.

In \cite{LN} a unifying picture to the problem was suggested on the basis of {\it degenerations}. The idea is to define, for a given amplitude $S_{i,f}$ associated with a given scattering process $i\rightarrow f$ an inclusive cross section formally defined as
\begin{equation}\label{LN}
\sum_{i'\in D(i),f'\in D(f)}|S_{i',f'}|^2 \komma 
\end{equation}
where $D(i)$ is the set of asymptotic states {\it degenerate} with the asymptotic state $i$. 

For the case of massless electrically charged particles the degeneration used in \cite{LN} for the case where the asymptotic state is a charged massless lepton with momentum $p$ is a state with the lepton having momentum $p-k$ and an additional on-shell photon with 4-momentum $k$ collinear to $p$.\footnote{For a recent discussion of the KLN theorem for QED see \cite{mcmullan} and references therein.}

At each order in perturbation theory the KLN recipe demands us to sum over all contributions at the same order in perturbation theory that we can build using degenerate incoming and /or outgoing states. Among these are specially interesting the {\it quantum interference effects} with disconnected diagrams where the additional photon entering into the definition of a degenerate incoming state is not interacting. In particular these interference effects play a crucial role to cancel collinear divergences in processes where the incoming electron emits a collinear photon, see for instance \cite{LN,mcmullan,Schwartz}.

The main target of this paper is to study the quantum consistency of the KLN prescription to define a quantum field theory of $U(1)$ massless charged particles. Our findings can be summarize in two main claims. First of all we provide substantial evidence about the existence of a KLN anomaly in the forward limit. This anomaly is worked out up to order $e^6$ in perturbation theory in appendix \ref{appendixC} and \ref{appendixD} explicitly. Secondly we argue that canceling this KLN anomaly is only posible if there exists a non-vanishing gauge anomaly to the $U(1)$ gauge theory. On one side we shall argue that the consistency of the KLN prescription in the forward regime implies the existence of a non vanishing gauge anomaly for the $U(1)$ gauge theory. This rules out the possibility of the existence of a theory with massless charges and long range interactions satisfying both KLN cancellation and being anomaly free. Secondly combining the {\it weak gravity conjecture} \cite{Nima} and {\it anomalous thresholds} for form factors, we derive a gravitational lower bound on the mass of the lighter massless charged fermion and a qualitative upper bound on the total number of fermionic species with the same charge as the electron.

\section{The KLN-theorem: degeneracies and energy dressing}

Let us briefly review the key aspect of the KLN theorem \cite{LN}. In order to do that let us consider scattering theory for a given Hamiltonian $H=H_0+gH_{I}$ and let us assume the Hamiltonian depends on a parameter $m$. Assuming a well defined scattering theory, the hamiltonian $H$ can be diagonalized using the corresponding M{\o}ller operators $U$. Let us denote $E_i(g,m)$ the corresponding eigenvalues. If for some value  $m_c$ of the mass parameter we have degenerations i.e.~$E_i(g,m_c) = E_j(g,m_c)$ then the perturbative expansion of $U_{i,j}$ becomes singular at each order in perturbation theory. However {\it at the same order} in perturbation theory the quantity $\sum_a U_{a,i}U^{*}_{j,a}$ where we sum over the set of states degenerate with the state $a$ is free of singularities in the limit $m=m_c$ leading to the prescription (\ref{LN}) for finite cross sections. The former result is true provided $\Delta_{a}( g,m)= (H_0-E)_{aa}$ has a good finite limit for $m=m_c$.

The quantum field theory meaning of $\Delta$ is the difference of energy between the bare and the dressed state. The theorem works if for fixed and finite UV cutoff the limit of this dressing effect is finite in the degeneration limit.

For the case of QED and for $m$ the mass of the electron, degeneracies appear in the limit $m_e\rightarrow0$. 
As stressed in \cite {LN} in this case the limit of $\Delta$ for $m_e\rightarrow0$ and fixed UV cutoff is not finite. The problem is associated with the well known behavior of the renormalization constant $Z$ for the photon field which goes as
\begin{equation}\label{Landau}
Z=1 - \frac{e^2}{6\pi^2} \log{\frac{\Lambda}{m_e}} \punkt
\end{equation}
The origin of the problem is well understood. Using K\"allen-Lehmann-representation to extract the value of $Z$ from the imaginary part of the bubble amplitude i.e. $\text{Im} D(k^2)$ for the photon propagator $D(k^2)$, we get for massless electrons a branch cut singularity in the physical sheet   for the threshold $k^2=0$ where the on-shell photon can go into a  collinear pair of on-shell electron and positron.

In \cite{LN} this problem was explicitly addressed and the suggested solution was to keep $m_e=0$ but to add a mass scale in the definition of $Z$ (see \cite{vaks}) associated with some IR resolution scale let us say $\delta$. The logic of this argument is to assume an IR correction of \eqref{Landau} where effectively $m_e$ is replaced by $\delta$ and to use this corrected $Z$  to define a  $\Delta$ non singular in the limit $m_e\rightarrow0$.

Note that the singular limit of $Z$ in the massless limit is the IR version of the famous Landau pole problem for QED. In this case we are not considering the limit where we send the UV cutoff to infinity but instead the limit $m_e\rightarrow 0$. In the massless limit there are contributions to the K\"allen-Lehmann-function coming from processes in which the on-shell photon with energy $\omega$ produces a pair of electron positron both collinear and on-shell. 
Incidentally note that in principle we have contributions of amplitudes where the on-shell photon decays into a set of a large number $n$ of electron-positron-pairs and photons where all of them are on-shell and collinear. The approach of the KLN program is to assume that after taking all these IR contributions into account the resulting $Z$, for fixed UV cutoff, is finite in the limit $m_e\rightarrow 0$. This does not imply solving the UV problem or avoiding the standard Landau pole, that depends on the sign in \eqref{Landau} and  that now will become dependent on the added resolution scale $\delta$. It simply means that for fixed UV cutoff the limit $m_e\rightarrow 0$ {\it could be} non singular. In section IV we will revisit the consistency of the limit $m_e\rightarrow0$ from a different point of view. 

Can we check the consistency of the KLN proposal perturbatively? To the best of our knowledge the KLN program of finding a redefinition of $Z$ where the cancellation can be defined in an effective way has not been developed. Thus we should expect that perturbative violations of the KLN theorem could appear whenever we work in the kinematical regime where originally appears the singularity responsible for the former behaviour of $Z$, namely in the forward regime $q^2=0$. 

In \cite{Schwartz} the authors presented a different but equivalent procedure to cancel infrared and collinear (IRC) divergences, where one has to sum either over degenerate initial {\it or} final states. Briefly, the cut-method defines IRC finite S-matrices by cutting the IRC divergent amplitude square and identify then new amplitude squares at the same order in perturbation theory. After summing over all these cutted diagrams the IRC divergences cancel each other. Within this set of new amplitude squares there occur diagrams that are interference terms of purely disconnected and thus forward scattered particles with diagrams that ensure to have the correct order in coupling constant, most properly loop diagrams. Some of these interference terms are IRC divergent and hence contribute to the cancelation of the IRC divergences and some of them are IRC finite and therefore not contributing to the cancelation scheme. A key difference however is that in the forward scattering process we are looking at we have an additional constraint for the outgoing photon momenta, namely $q^\mu+k^\mu=k'^\mu$. This constraint ensures, once integrate over the photon momentum, that we will only get a single $\log(m_e)$ divergence.

\section{Degeneracies and anomalous thresholds}
\begin{figure}[t]
	\centering
	\includegraphics[width=0.4 \textwidth]{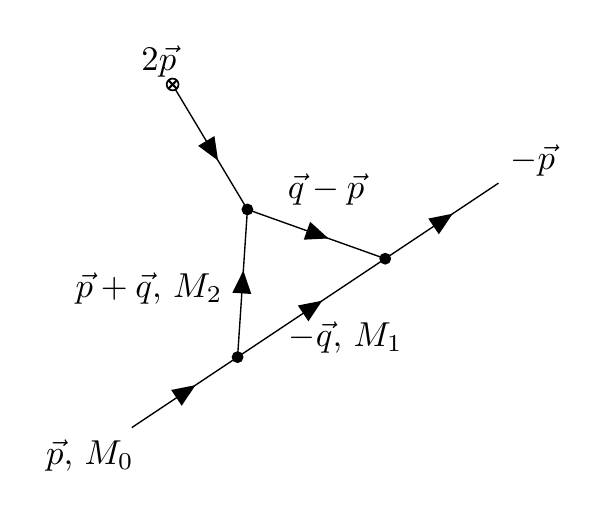}
	\caption{Anomalous threshold in Breit frame.}	\label{threshold}
\end{figure}
In scattering theory the existence of {\it anomalous thresholds} for form factors of bound states is well known (see \cite{Eden,mandelstam,Cutkosky}). The idea is simply to consider the triangular contribution to the form factor of a particle $A$ by some external potential. If the particle $A$ can decay into a pair of particles $N$ and $B$ where only $N$ interacts with the external potential  we get the triangular amplitude depicted in figure \ref{threshold}. If we now impose all the internal lines to be on-shell we can find a critical transfer momentum for which the corresponding amplitude has a leading Landau singularity in the physical sheet. This transfer momentum defines the anomalous threshold. This leads to a logarithmic contribution to the amplitude and to a non vanishing absorptive part forbidden by standard unitarity. The simplest way to set when this singularity is physical is using the Coleman-Grossman-theorem \cite{Coleman} that dictates that the singularity is physical if the triangular diagram can be interpreted as a space-time physical process with energy momentum conservation in all vertices and with the internal lines on-shell i.e. as a Landau-Cutkosky-diagram. 

\begin{figure}[t]
	\centering
	\includegraphics[width=0.4 \textwidth]{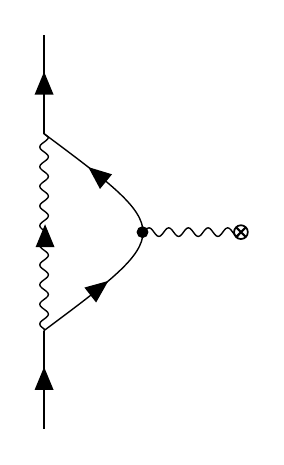}
	\caption{Landau Cutkosky diagram associated with the anomaly.}	\label{triangle2}
\end{figure}

Let us now consider the degenerations as formally representing the massless electron as a composite state of electron and collinear photon. In this case we can consider the triangular contribution in figure \ref{triangle2} to the form factor where the electron in the triangle interacts with the external potential. In this case it is easy to see that an anomalous threshold can appear only in the forward limit when the transfer momentum $q^2$ is zero (see Appendix A).

From the KLN theorem point of view we can associate these kinematical conditions to the degeneration defined by the absorption and emission process of a collinear photon with the same value of the 4-momentum $k$ and with $k$ collinear to $q$. In this case the logarithmic divergence $\log(m_e)$ of the anomalous threshold can be canceled with the corresponding KLN sum. 

However the KLN prescription in this forward limit allows us to have different 4-momentum $k$ and $k'$ for the absorbed and emitted photon. If this amplitude is logarithmically divergent it cannot be trivially canceled by a one loop contribution to the form factor. Next we shall see that this is indeed the case and that the only possible cancellation leading to a consistent theory of massless charged particles is using quantum interference with processes controlled by the triangular graph defining the gauge anomaly of the underlying gauge theory.
 
\section{The KLN anomaly}
In this section we shall consider the absorption emission process in the forward limit with $k\neq k'$. Let us fix as data of the form factor scattering process the 4-momentum of the initial electron $p$ and the exchanged energy-momentum that will denote $q$. Let us denote the amplitude $S(p,q)$. For these data the KLN prescription requires to define the sum
\begin{equation}\label{LN2}
\sum_{n_i,n_f} |S(p,q; n_i,n_f)|^2 \komma
\end{equation}
where $n_i$ and $n_f$ denote the different degenerate states contributing to the process that are characterized by the number $n_i$ of absorbed collinear photons attached to the incoming line and the number $n_f$ of emitted collinear photons attached to the outgoing line. All of these photons are assumed to have energies bigger than the IR energy resolution scale set by the Bloch-Nordiesk-recipe. Generically each term in the sum \eqref{LN2} involves the integral over the 3-momentum of the collinear photons within a given angular resolution scale. The amplitudes in (\ref{LN2}) contain internal lines with the corresponding propagators being on-shell. 

In what follows we shall be interested in the {\it forward corner} of phase space characterized by vanishing transfer momentum, i.e.~
\begin{equation} \label{transferemomentum}
q^2 =0 \punkt
\end{equation}
In the forward regime the first absorption emission process contributing to the sum contains one absorbed  photon and one emitted photon. This process is characterized by the following set of kinematical conditions $ p q\approx p k    \approx  p' k'   \approx 0$. This implies that in this corner of phase space the two propagators entering into the amplitude are on-shell. This after integration leads to a collinear divergence. Moreover in these kinematical conditions we have 
\begin{equation}
 k' - k =  q
\end{equation}
and, as mentioned, in the forward limit the outgoing electron has the same momentum as the in-coming one, i.e.~$ p=  p'$. 
Since for this amplitude both the absorbed and the emitted photons are collinear to the incoming and outgoing electron respectively, the KLN recipe indicates that this divergence should be canceled by the collinear contribution of virtual photons running in the loop. 

\begin{figure}[b]
	\centering
	\includegraphics[width=0.5 \textwidth]{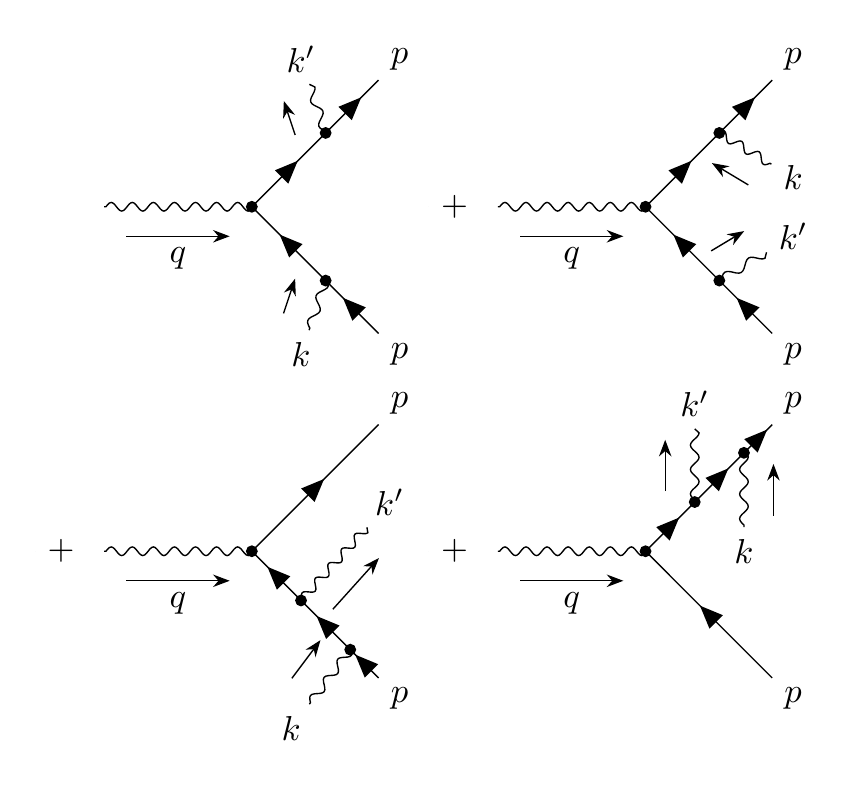}
	\caption{These diagrams lead to a mass divergence once \eqref{LN} is applied, with the kinematics $k'=k+q$.}	\label{figure1}
\end{figure}

In what follows we shall show that in the forward limit emission absorption processes with $k\neq k'$ lead to logarithmic divergences. The diagrams that lead to the collinear term are given in figure \ref{figure1}. We work in the chiral basis and choose the kinematics for the electron to run in z-direction. In the appendix \ref{appendixB} we explain the details and the notations used in the calculation. We omit all terms that will not lead to a collinear divergence. In these conditions we get for the amplitudes for a forward scattered right-/left-handed elctron
\begin{widetext}\begin{align}\label{amp}
	\im M^R = &- \im e^3\frac{\sqrt{2}\theta\left[ \omega(\omega+\omegaq)+(2E+\omega\lambda)(2E+(\omega+\omegaq)\lambdas) \right]}{E\omega\omegaq \left( \theta^2+\frac{m^2}{E^2} \right)} \nonumber \\ &+\im e^3 \frac{\sqrt{2}\theta\left[-\omega\omegaq+(2E+\omega\lambda)(2E+\omegaq\lambdaq)\right]}{E\omega(\omega+\omegaq)\left(\theta^2+\frac{m^2}{E^2}\right)} \komma \\   
	\im M^L = &- \im e^3\frac{\sqrt{2}\theta\left[ \omega(\omega+\omegaq)+(2E-\omega\lambda)(2E-(\omega+\omegaq)\lambdas) \right]}{E\omega\omegaq \left( \theta^2+\frac{m^2}{E^2} \right)} \nonumber \\ &+\im e^3 \frac{\sqrt{2}\theta\left[-\omega\omegaq+(2E-\omega\lambda)(2E-\omegaq\lambdaq)\right]}{E\omega(\omega+\omegaq)\left(\theta^2+\frac{m^2}{E^2}\right)} \punkt
\end{align}\end{widetext}
In order to use the KLN-theorem we need to perform the integration over photon momenta $\int\frac{\diff^3\vec k}{(2\pi)^3 2\omega}$, 
and taking into account the constraint $k'=q+k$, coming from the conservation of energy and momentum. The interesting part of the integral is the one over the small angle $\theta$, since there the collinear divergence shows up. In the collinear limit  $\omegas=\omegaq+\omega$ and $\thetas= \frac{\omegaq \thetaq}{\omega+\omegaq}$ (see appendix \ref{appendixB}). 
Including these constraints, and integrating over the phase space $\int\frac{\diff^3\vec k}{(2\pi)^3 2\omega}$ with small angle $\theta$ gives
\begin{widetext}\begin{gather}
	\int\frac{\diff^3\vec k}{(2\pi)^3 2\omega} \frac{1}{4}\sum_\text{spins}\abs{\im M}^2   = \nonumber \\
	 \int\frac{\diff\omega\,\omega}{(2\pi)^2} \frac{e^6}{4E^2\omega^2}\logEdeltam \left[ \frac{-\omega\omegaq+(2E+\omega\lambda)(2E+\omegaq\lambdaq)}{(\omega+\omegaq)} -\frac{ \omega(\omega+\omegaq)+(2E+\omega\lambda)(2E+(\omega+\omegaq)\lambdas) }{\omegaq }   \right]^2 \nonumber \\
	+\left( \lambda\rightarrow-\lambda, \, \lambdas\rightarrow-\lambdas, \, \lambdaq\rightarrow-\lambdaq\right) \label{ampsquare} \komma
\end{gather}\end{widetext}
where $\delta$ is a small angular resolution scale. The details of the calculations can be seen in the appendix \ref{appendixB}.

In summary for generic $q$ and for emission absorption processes we get a double pole for $k=k'$ that can interfere with a disconnected diagram where the photon is not interacting. For $q^2=0$ we have a double pole on the kinematical sub manifold defined by $k-k'=q$ that leads, for fixed $q$ and after integration over $k$, to a collinear divergence that don't interfere with disconnected diagrams where the photon is not interacting.
Thus we have obtained an additional collinear divergent contribution from the KLN-theorem \eqref{LN}, which is not canceled by any known loop factors. We will refer to this contributions as a KLN anomaly. 

\section{The KLN anomaly and the triangular anomaly}\label{KLNanomaly}
From a perturbative point of view a crucial ingredient of anomalies in four dimensions are triangle Feynman diagrams with currents inserted at the vertices. This is the case for the original ABJ anomaly \cite{Adl,BJ} as well as for gauge anomalies. The difference lies in the type of currents we insert in the vertices.

The analytic properties of triangular graph amplitudes were extensively studied in the early 60's using Landau equations \cite{Landau-ana,CN} and Cutkosky rules \cite{Cut}. As already mentioned it was first observed in \cite{Nambu} the existence, for triangular graphs, of singularities associated with non unitary cuts. These singularities are the {\it anomalous thresholds} \cite{Cutkosky} (see Appendix \ref{anomalousthreshold} for the relevant formulae).

In reference \cite{DZ} it was first pointed out the connection of the anomaly with the IR singularities of the corresponding triangular graph amplitude. This approach was further developed in \cite{Fris} and \cite{Coleman} in the context of t'Hooft's anomaly matching conditions \cite{tHooft}. 
\begin{figure}[t]
	\centering
	\includegraphics[width=0.4 \textwidth]{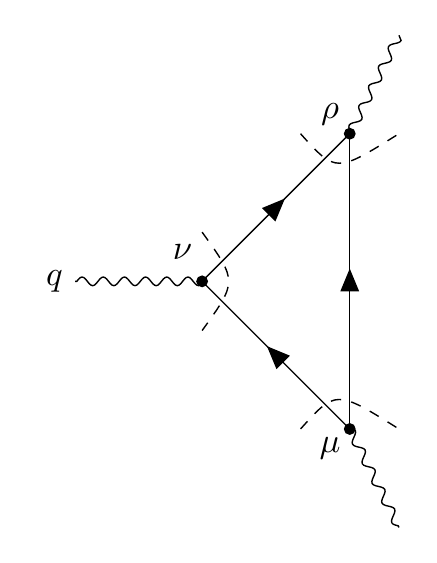}
	\caption{The anomaly diagram with the non unitary cuts.}	\label{cutting}
\end{figure}

Let us first briefly recall the analytic structure of anomalies. In a nutshell given a triangular amplitude $\Gamma^{\mu\nu\rho}$ for three chiral currents let us denote $\Gamma(q^2)$ the invariant part of the amplitude for $q^2$ the relevant transfer momentum (see figure \ref{cutting}). The anomaly is defined as the residue of $\Gamma(q^2)$ at $q^2=0$, i.e.
\begin{equation}\label{anomaly}
q^2\Gamma(q^2) = \cal{A} \komma
\end{equation}
for $\cal{A}$ the c-number setting the anomaly. Standard dispersion relations
connect (\ref{anomaly}) with the imaginary part of $\Gamma(q^2)$, namely
\begin{equation}
\text{Im} \Gamma(q^2) \sim \delta(q^2) \punkt
\end{equation}
The physical meaning of the singularity underlying the anomaly requires to understand the analytic properties of the full amplitude.

As already mentioned for the triangular graph we can have normal threshold singularities as well as the anomalous threshold singularities that correspond to the leading Landau singularity. In the language of Landau equations the normal threshold corresponds to the reduced graph where the Feynman parameter $\alpha$ of one of the three lines is equal to zero. In what follows we shall discuss the anomalous threshold.\footnote{Normal thresholds are relevant for the study of chiral anomalies in two dimensions. In this case the leading singularity for the corresponding two point diagram represents the $\eta'$ \cite{Witten}.} This corresponds to put the three lines of the triangle on-shell. The threshold is determined by the value of transfer momentum $q^2$ at which the corresponding diagram with all the internal lines on-shell and with external real photons is kinematically allowed. For massless particles running in the triangle this anomalous threshold exists and it is given by $q^2 =0$. The corresponding discontinuity is determined by Cutkosky rules as 
\begin{equation}\label{disc}
\int \diff^4p \prod \theta(p_i^0)\delta(p_i^2)\prod C_i \komma
\end{equation}
where the $C_i$ are the {\it physical values} of the three amplitudes determined by the non unitary cut (see figure \ref{cutting}).\footnote{In (\ref{disc}) we have formally included in the $C_i$ the propagator factors distinguishing bosons from fermions in the cuted lines.} As shown in \cite{Coleman} the discontinuity of the triangular amplitude goes as
\begin{equation}
q\delta(q^2)
\end{equation}
and it is non vanishing.
Let us now look at this discontinuity as an anomalous threshold. The physical process associated with this discontinuity can be understood as a real incoming photon that for massless charges decays into a pair of collinear on-shell electron and positron. One piece of the pair interacts with the external potential with some transfer momentum and finally the pair annihilates giving rise to a massless photon. Note that the discontinuity for the anomaly graph relies on the fact that for massless charges the photon can decay into a pair of on-shell collinear charged particles. If we fix one chirality for the running electron this discontinuity gives us the anomaly. To cancel the gauge anomaly for $U(1)$ we need to have real representations i.e.~to add both chiralities in the loop.

The decay of the photon into a pair of collinear massless fermions can be formally interpreted as a degeneracy between the photon and a pair of collinear massless charged particles. From this point of view the anomaly is just the anomalous threshold associated with this formal compositeness of the photon. In more precise terms what makes the anomaly {\it anomalous} is {\it the existence of an absorptive part of the triangular amplitude that is expected, from standard unitarity (only one cut), to vanish.}\footnote{The anomaly matching \cite{tHooft} reflects that the discontinuity of the triangular graph is the same for the IR and UV physical spectrum running in the triangle.}

Let us now relate the KLN anomaly and the triangular anomaly. As discussed the KLN anomaly appears whenever $k\neq k'$ with zero transfer momentum \eqref{transferemomentum}. From the KLN theorem point of view the contribution computed in the former section should cancel with some contribution to the form factor of the electron. 

Since we are working at order $e^6$ we need, in principle,  to include all loop diagrams to this order in perturbation theory contributing to the form factor. The interference term of two-loop diagrams and the tree-level diagram and the interference term of a one-loop diagram with one incoming collinear photon and a tree-level diagram with also one incoming collinear photon are of order $e^6$. We treat these diagrams and its collinear divergent contribution to the amplitude in the appendix \ref{appendixC} and \ref{appendixD}. The two-loop contribution treated in \ref{appendixD} goes like $\log^2 (m_e) $ and therefore can not cancel the KLN anomaly. The interference term treated in \ref{appendixC} is of order $\log(m_e)$ but will not cancel the KLN anomaly as shown in appendix \ref{appendixC}. Thus, the $\log$-divergent term in the amplitude square \eqref{ampsquare} can't be canceled. This is intuitively clear from the fact that the KLN anomaly appears when $k\neq k'$.

\begin{figure}[t]
	\centering
	\includegraphics[width=0.4 \textwidth]{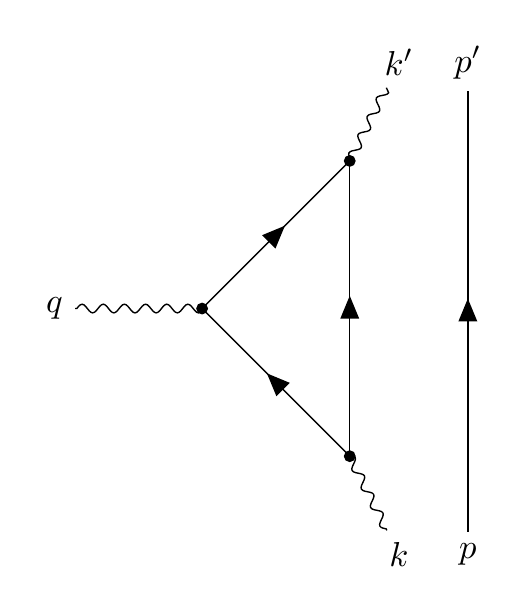}
	\caption{Anomaly triangle diagram with disconnected electron contributing to the amplitude square.}	\label{triangle1}
\end{figure}

However, in this case we have the possibility of defining an interference term at this order in perturbation theory. Namely, we can think a diagram where we have the electron non interacting and where the companion collinear photon is interacting {\it through the triangular anomaly} with the external source. This allows $k\neq k'$ in the forward limit where $k$ and $k'$ are both collinear to the momentum $p$ of the electron. The role of the triangular anomaly graph is to account for the difference between $k$ and $k'$ and to provide the needed logarithmic singularity. Thus
for $k\neq k'$ the only possible contribution will come from the interference with the anomaly diagram in figure \ref{triangle1}. 
Therefore for fixed chirality of the electron in figure \ref{figure1} the only possibility to cancel the KLN anomaly is to assume a non vanishing value for the triangular graph. However, this is only possible if the corresponding gauge theory is anomalous. In fact once we sum over all chiralities in the triangle we get a zero contribution to a form factor with $k\neq k'$. In summary we have shown that

{\it The KLN anomaly can be only canceled if the gauge theory is anomalous.}

Consequently we conclude that the the KLN anomaly can be only canceled effectively adding a mass for the charged fermions.

In summary the KLN anomaly in the forward limit with $k\neq k'$ corresponds to an anomalous threshold in the form factor where it is the photon, the one that interacts with the external potential. This can only take place through the triangular graph and it is only non vanishing if the theory is anomalous with respect to the underlying gauge symmetry.

Before finishing this section let us make two brief comments that could help to clarify the argument. First of all note that in \cite{LN} processes as the ones in figure \ref{figure1} were considered. For generic $q$ this produces logarithmic divergences only in the case $k=k'$ and these are compensated using a disconnected diagram where the companion photons is not interacting. In the particular case of $q^2=0$ we have a collinear divergence even for $k\neq k'$ and the corresponding disconnected diagram is now the one in figure \ref{triangle1} where we need to include the triangular anomaly in the photon line. The second comment concerns the recent discussion of symmetries in massless QED \cite{NewSymmetries}. The first thing to be noticed is that in the collinear case the corresponding {\it dressing} using coherent states \cite{FK} is ill defined (see \cite{us} for a brief discussion). In the symmetry language this could be interpreted as indicating that KLN recipe is violating these symmetries. Actually a potential way to interpret our result is that in the massless case the collinear dressing in the forward limit $q^2=0$ is actually incompatible with non anomalous gauge invariance.\footnote{In \cite{Gia} it is argued that non vanishing gravitational topological susceptibility implies the absence of massless fermions.}

In case the origin of the transfer momentum is gravitational the situation is more interesting and richer. In fact in this case although we keep the same electromagnetic degeneration due to collinear electromagnetic radiation the external field, once it is assumed to be gravitational, can contribute to the form factor due to the graviton photon vertex. The analysis of this case is postponed to a future work.

\section{A lower bound for the electron mass}
In the former section we have argued that a quantum theory of massless charged fermions is inconsistent. The core of the argument is that consistency requires to cancel the KLN anomaly and that is only possible if the theory has non vanishing $U(1)$ gauge anomaly i.e.~if the theory is by itself inconsistent.

In what follows we shall put forward the following conjecture:

{\it In a theory with minimal length scale $L$ the minimal mass of a $U(1)$ charged fermion, for instance the electron, is given by
\begin{equation}\label{conj}
m_e \geq \frac{\hbar}{L} \text{e}^{-\frac{1}{e^2 \nu}} \komma
\end{equation}
where $e^2$ is the corresponding coupling and $\nu$ is the number of fermionic species with charge equal to the electron charge.}

Before sketching the argument let us make explicit the logic underlying this conjecture. The bound \eqref{conj} can be naively obtained from the perturbative expression \eqref{Landau} as the minimal mass of the electron consistent with pushing the perturbative Landau pole to $\frac{\hbar}{L}$. To argue in that way will force us to assume that the perturbative result for $Z$ already rules out the consistency of a theory of massless electrons. This will contradict the basic assumption of the KLN theorem of the potential redefinition of $Z$ with a well defined $m_e\rightarrow0$ limit. Thus our approach to set a bound on the electron mass will consist in looking for some anomalous threshold singularity depending on the electron mass and to set the bound by analyzing the limit $m_e\rightarrow0$ of these contributions to form factors.

In order to look for the appropriated form factor we shall use the constraints on the charged spectrum coming from the weak gravity conjecture \cite{Nima}. This conjecture is equivalent to say that in absence of SUSY extremal electrically charged black holes are unstable. This leads to the existence in the spectrum of a particle with mass satisfying
\begin{equation}\label{WGC}
m^2_e \leq e^2 M_P^2 \punkt
\end{equation}
Once we accept the instability of charged black holes in absence of SUSY we can compute the effect {\it of this instability to the form factor of the charged black hole in the presence of an external electric potential}. Denoting $m_e$ the mass of the minimally charged particle we have again the anomalous threshold contribution where the black hole interaction with the external potential is mediated by the charged particle through the corresponding triangular graph. Assuming all the particles in the process {\it to be on-shell} the anomalous threshold is given by
\begin{equation}
t_0 = 4m_{e}^2 - \frac{\left(M_{bh}^2 - {M'}_{bh}^2 -m_e^2\right)^2}{{M'}_{bh}^2} \komma
\end{equation}
where we think the instability as the decay of a black hole  of mass $M$ and charge $Q$ into a smaller black hole of mass $M'$ and a particle with mass $m$ and  minimal charge $e$ that we will call the electron (see figure \ref{BH}). 

As shown in appendix A for the typical gravitational binding energy that we expect for a black hole the anomalous threshold contribution to the corresponding amplitude will go as
\begin{equation}
\log\left(\frac{M_P}{m_e}\right) \komma
\end{equation}
where we have used as UV cutoff the Planck mass.

\begin{figure}[t]
	\centering
	\includegraphics[width=0.4 \textwidth]{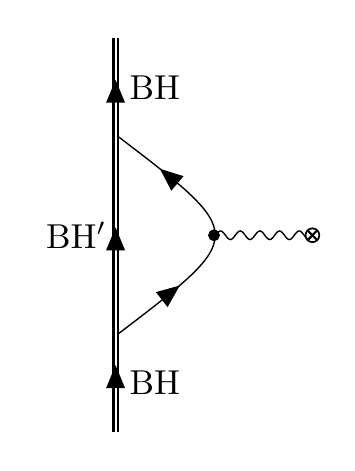}
	\caption{Anomalous threshold for the form factor of a RN black hole.}	\label{BH}
\end{figure}
The imaginary part of this amplitude can be interpreted as an anomalous threshold to the absorptive part of the form factor of the charged black hole in the presence of an external electromagnetic field. Now we have what we were looking for, namely a {\it physical amplitude} that depends on the electron mass in a way that is singular in the massless limit. In order to avoid the singular limit $m_e\rightarrow0$ we can impose, on the basis of unitarity, that the corresponding amplitude is smaller than one. If we do that we get
\begin{equation}
\nu C^2\log\left(\frac{M_P}{m_e}\right) \leq 1 \komma
\end{equation}
where $C$ represents the physical decay amplitude of the black hole to emit an electron. If we assume this amplitude to be proportional to the electromagnetic coupling we get
the lower bound above. Here $\nu$ is the number of charged fermionic species with equal charge to the electron.\footnote{The role of electrically charged species is analogous to the one suggested originally by Landau \cite{LandauLondon} to lower the Landau pole.} Taking seriously the former bound on the electron mass leads to an upper bound on the number of fermionic species with the electron charge of the order of $11$ species.\footnote{Adding the effect of gravitational species \cite{Giaspecies} will multiply the former bound by a global suppression factor $\frac{1}{\sqrt{N_g}}$.} The key point to be stressed here is that in deriving this bound we don't use the perturbative Landau pole but instead the anomalous threshold singularity we get assuming the gravitational instability of RN extremal black holes.

A different way to understand the anomalous threshold is as follows. For the case of standard black holes with entropy $N$ we should expect that the threshold for an absorptive part should be $t_0 \sim O(1/N)$ in Planck units i.e.~absorption of one information bit. The existence of massless charged particles pushes down this threshold to the anomalous value $O(m_e^2)$ and therefore we could expect a lower {\it information bound} for the mass of the electron $m_e \sim 1/N$ in Planck units for the largest possible black hole. Thus and using a cosmological bound for the largest black hole we could conclude that the lower bound on the mass of electrically charged fermions is given, in Planck units, by $\frac{1}{\sqrt{N_{H}}}$ with $N_{H}$ determined by the Hubble radius of the Universe as $\frac{R_{H}^2}{L_P^2}$. 

To end let us make a comment on \eqref{WGC}. For equality this can be written as $e^2 = \frac{m_e^2}{M_P^2}$. Thinking in a diagram representing an energetic Planckian photon decaying into a set of $n$ on-shell pairs and estimating $n \sim \frac{M_P}{m_e}$ the former relation \eqref{WGC} simply express the {\it criticality condition} \cite{Dvali:2012en} $e^2 \sim \frac{1}{n}$ typical of classicalization. 

Before ending we would like to make a very general comment on black hole physics intimately related with the former discussion. In \cite{portrait} we put forward a {\it constituent portrait} of black holes. The most obvious consequence of this model is the prediction of anomalous thresholds in the corresponding form factors at small angle. On the other hand these anomalous thresholds define a canonical example of in principle observable {\it quantum hair}.

\section{Final comment}
It looks like that nature abhors massless charged particles whenever the charge is associated with a long range force as electromagnetism. This is not a serious problem for confined particles but it is certainly a problem for charged leptons. Taken seriously, it will means that the limit with vanishing Yukawa couplings should be quantum mechanically inconsistent. In string theory we count with a geometrical interpretation of Yukawa couplings in terms of intersections \cite{Yukawa} and in some constructions based on brane configurations in terms of world sheet instanton contributions. It looks like that a consistency criteria for string compactifications should prevent the possibility of massless charged leptons and consequently of vanishing Yukawa couplings. The problem of a consistent massless limit of leptons is on the other hand related with the problem of {\it naturalness} in t'Hooft's sense \cite{tHooft}. Naively the symmetry enhancement that will make natural the massless limit is chiral symmetry. What we have observed in this note can be read from this point of view. The IR collinear divergences, if canceled in the way suggested by the KLN-theorem, prevent the realization of this chiral symmetry indicating the unnatural condition of the massless limit of charged leptons. A hint in that direction was the observation of \cite{LN} about the existence for massless QED of non vanishing helicity changing amplitudes  in the absence of any supporting instanton like topology. Thus, it looks that the existence of a fundamental lower bound on the mass of charged leptons is inescapable.

\appendix
\section{Anomalous threshold kinematics} \label{anomalousthreshold}
Let us consider the leading Landau singularity for the diagram in figure \ref{threshold}. This corresponds to have all the internal lines of the diagram on-shell satisfying energy momentum conservation in the three vertices.
Following \cite{Cutkosky} the diagram is presented in Breit frame. The transfer momentum is given by $-4\vec{p}^2$, the normal threshold is given by $4M_2^2$ where $M_2$ is the mass of the particle in the triangle interacting with the external source. The anomalous threshold associated with the leading Landau singularity is given by
\begin{equation}
	t_0 = 4M_2^2-\frac{M_0^2-M_1^2-M_2^2}{M_1^2} \komma
\end{equation}
where $t_0=-4\vec{p}^2_{0}$. This is the minimum momentum where all the particles in figure \ref{threshold} can be on-shell. Here also the scattering angles have to be below a small threshold which in our case refer to the resolution scale angle $\delta$. Note that this anomalous threshold is independent on the energy of the process. The reason for calling it anomalous is that it is smaller than the normal threshold given by standard unitarity.

As discussed in the text the discontinuity associated with this singularity can be computed using the Cutkosky rules for the diagram. The corresponding amplitude contains a term proportional to $\log\left(1-\frac{t}{t_0}\right)$. For the diagram in figure \ref{triangle2} where we use the degeneration between the electron and a pair electron and collinear photon (both on-shell) the anomalous threshold gives the $\log(m_e)$ terms in the amplitude. 

In order to get a clearer picture of the underlying kinematics we can compute the relative velocity $v$ between the two particles $1$ and $2$. This is given by the so called K\"allen-function
\begin{equation}
v = A(M_0^2 -M_1^2-M_2^2) \komma
\end{equation}
with $A^2= M_0^4+M_1^4+ M_2^4 -2M_0^2M_1^2 -2M_0^2M_2^2-2M_1^2M_2^2$. In the degenerate case with $M_0= M_2= m_e$ and $M_1 =m_{\gamma}$ the mass of a photon we get the limit $v= i\infty$ corresponding to particles $1$ and $2$ moving collinearly i.e.~they remain coincident.

Introducing a {\it binding energy} as $M_0+B = M_1+M_2$ we observe that for $M_0<M_1+M_2$ the velocity $u$ defined above is imaginary reaching collinearity in the limit $B\rightarrow 0$. Moreover in the limit where $M_1$ is much larger than $M_2$ the anomalous threshold can be approximated by:
\begin{gather}
	t_0 \approx 4 B m_e \left( 2 - \frac{B}{m_e} \right) \punkt
\end{gather}
In the gravitational case $t_0$ goes from zero in the limit $B\rightarrow 0$ to the normal threshold $4m_e^2$ in the limit of maximal gravitational binding energy.

\section{Notation and calculation for the amplitudes} \label{appendixB}
We set the kinematics of the forward scattered right- or left-handed electron in such way that the electrons runs with momentum $\abs{\vec p} $ along the z-axes, i.e.~for the 4-momentum of the electron we have
\begin{gather}
	p^\mu=\begin{pmatrix} E \\ 0 \\ 0 \\ \abs{\vec p} \end{pmatrix} \komma
\end{gather}
where $E$ is the energy of the electron. The Dirac spinor for the right-/left-handed electron in chiral representation is given by
\begin{gather}\label{diracspinors}
	u^R(p)  := \begin{pmatrix} 0 \\ u_R \end{pmatrix} \text{ and } u^L(p)  := \begin{pmatrix} u_L \\ 0 \end{pmatrix} \komma 
\end{gather}
In the limit where the mass $m_e$ of the electron goes to $0$
\begin{gather}\label{spinors}
	u_L = \sqrt{2E} \begin{pmatrix}0 \\1 \end{pmatrix} \text{ and } u_R = \sqrt{2E} \begin{pmatrix} 1\\0 \end{pmatrix} \komma 
\end{gather}	
and
\begin{gather}
	 p^\mu \approx \begin{pmatrix} E\left(1+\frac{m_e^2}{2E^2}\right) \\ 0 \\ 0 \\ E \end{pmatrix}\komma \label{papprox}
\end{gather}
holds. We work in Weyl (chiral) basis where the $\gamma$-matrices are given by
\begin{gather}\label{weyl}
	\gamma^\mu = \begin{pmatrix} 0 & \sigma^\mu \\\bar{\sigma}^\mu &0   \end{pmatrix} \komma
\end{gather}
with $\sigma^\mu = (1,\sigma^i)$ and $\bar {\sigma}^\mu = (1,- \sigma^i)$, where $\sigma^i$ are the standard Pauli matrices. 

We start with the first amplitude of the diagrams in figure \ref{figure1}. The notation will be $\im M^R$/$\im M^L$  is the amplitude where the electron is right-/left-handed before and after the scattering. We keep the electrons helicity equal in the scattering process since we are not interested in helicity flipping processes. Then, writing the amplitudes in terms of 2x2 matrices figure \ref{figure1} gives the amplitudes
\begin{widetext}\begin{align}
	\im M_1^R = \frac{-\im e^3}{(2pk)(2pk')}&\left( u^\dagger_R \, \varepsilons^* \cdot \sigma \, (p+k') \cdot \bar \sigma \, \varepsilonq \cdot \sigma \, (p+k) \cdot \bar \sigma \, \varepsilon \cdot \sigma \, u_R \right) \label{amprawR1} \komma \\
	\im M_1^L = \frac{-\im e^3}{(2pk)(2pk')}&\left( u^\dagger_L  \, {\varepsilons}^* \cdot \bar \sigma \, (p+k') \cdot \sigma \, \varepsilonq \cdot \bar \sigma \, (p+k) \cdot \sigma \, \varepsilon \cdot \bar \sigma \, u_L\right) \label{amprawL1} \komma \\
	\im M_2^R = \frac{-\im e^3}{(2pk)(2pk')}&\left( u^\dagger_R \, \varepsilon \cdot \sigma \, (p-k) \cdot \bar \sigma \, \varepsilonq \cdot \sigma \, (p-k') \cdot \bar \sigma \, {\varepsilons}^* \cdot \sigma \, u_R \right) \label{amprawR2} \komma \\
	\im M_2^L = \frac{-\im e^3}{(2pk)(2pk')}&\left( u^\dagger_L  \, \varepsilon \cdot \bar \sigma \, (p-k) \cdot \sigma \, \varepsilonq \cdot \bar \sigma \, (p-k') \cdot \sigma \, {\varepsilons}^* \cdot \bar \sigma \, u_L\right) \label{amprawL2} \komma 
\end{align}
and
\begin{align}
	\im M_3^R = \frac{\im e^3}{(2pk)(2pq)}&\left( u^\dagger_R \, \varepsilonq \cdot \sigma \, (p-q) \cdot \bar \sigma \, {\varepsilons}^* \cdot \sigma \, (p+k) \cdot \bar \sigma \, \varepsilon \cdot \sigma \, u_R \right) \label{amprawR3} \komma \\
	\im M_3^L = \frac{\im e^3}{(2pk)(2pq)}&\left( u^\dagger_L  \, \varepsilonq \cdot \bar \sigma \, (p-q) \cdot \sigma \, {\varepsilons}^* \cdot \bar \sigma \, (p+k) \cdot \sigma \, \varepsilon \cdot \bar \sigma \, u_L\right) \label{amprawL3} \komma \\
	\im M_4^R = \frac{\im e^3}{(2pk)(2pq)}&\left( u^\dagger_R \, \varepsilon \cdot \sigma \, (p-k) \cdot \bar \sigma \, {\varepsilons}^* \cdot \sigma \, (p+q) \cdot \bar \sigma \, \varepsilonq \cdot \sigma \, u_R \right) \label{amprawR4} \komma \\
	\im M_4^L = \frac{\im e^3}{(2pk)(2pq)}&\left( u^\dagger_L  \, \varepsilon \cdot \bar \sigma \, (p-k) \cdot \sigma \, {\varepsilons}^* \cdot \bar \sigma \, (p+q) \cdot \sigma \, \varepsilonq \cdot \bar \sigma \, u_L\right) \label{amprawL4} \komma 
\end{align}\end{widetext}
where we omitted the terms proportional to the electron mass because they will give no collinear divergent term in the limit $m\rightarrow 0$ and $a\cdot b$ is the normal scalar product in 4d Minkowski space. The notation is $\varepsilon^\mu = \varepsilon^\mu(\lambda,\theta,\phi)$, ${\varepsilons}^\mu = \varepsilon^\mu(\lambda',\thetas,\phis)$, $\varepsilonq^\mu = \varepsilon^\mu(\lambda_q,\thetaq,\phiq)$ with 
\begin{gather}
	\varepsilon^\mu(\lambda,\theta,\phi) = \frac{1}{\sqrt{\cos^2(\theta)+1 }} \begin{pmatrix} 0 \\ \exp(-\im\lambda \phi)\cos(\theta)\\\im \lambda \exp(-\im \lambda \phi ) \cos(\theta) \\-\sin(\theta) \end{pmatrix}  \komma
\end{gather}
and $k^\mu = k^\mu(\omega,\theta,\phi)$, ${k'}^\mu = k^\mu(\omegas,\thetas,\phis)$, $q^\mu = k^\mu(\omegaq,\thetaq,\phiq)$ with 
\begin{gather}
	k^\mu(\omega,\theta,\phi)=\omega \begin{pmatrix} 1\\\sin(\theta)\cos(\phi) \\\sin(\theta)\sin(\phi) \\\cos(\theta) \end{pmatrix} \komma
\end{gather}
where $\omega$ is the energy, $\lambda$ the polarization and $\theta$ and $\phi$ the scattering angles of the corresponding photon. For example, a photon with polarization vector $\varepsilon^\mu$ and $\lambda=+1$/$-1$ is an incoming right-/left-handed photon.

In the collinear limit the angles $\theta$, $\thetas$ and $\thetaq$ appearing in the calculations are small, i.e.~$\cos\theta \approx 1-\frac{\theta^2}{2}$, $\cos\thetas \approx 1-\frac{\thetas^2}{2}$ and $\cos\thetaq \approx 1-\frac{\thetaq^2}{2}$. So that together with \eqref{papprox} we can approximate 
\begin{gather}
	2pk \approx E \omega \left( \frac{m^2_e}{E^2}+\theta^2 \right) \label{prop} \komma \\ 
	2pk' \approx E \omega \left( \frac{m^2_e}{E^2}+\thetas^2 \right) \label{props} \komma \\
	2pq \approx E \omega \left( \frac{m^2_e}{E^2}+\thetaq^2 \right) \label{propq} \punkt
\end{gather} 
A simple Taylor expansion to first order in $\theta$ and matrix multiplication shows that in general for the right-handed electron  
\begin{gather}
	(p\pm k(\omega,\theta,\phi))\cdot \bar \sigma \, \varepsilon(\lambda,\theta,\phi) \cdot \sigma\, u_R  \approx \sqrt{2}\theta\left(E\pm \frac{\omega}{2}(1+\lambda)\right) u_R  \label{LNidentity1} \komma \\ 
	u^\dagger_R  \, \varepsilon(\lambda,\theta,\phi) \cdot  \sigma \, (p\pm k(\omega,\theta,\phi)) \cdot \bar \sigma \approx \sqrt{2}\theta\left(E\pm \frac{\omega}{2}(1-\lambda)\right) u^\dagger_R \komma
\end{gather}
holds and for the left-handed electron
\begin{gather}
	(p\pm k(\omega,\theta,\phi)) \cdot \sigma \, \varepsilon(\lambda,\theta,\phi) \cdot \bar \sigma \, u_L  \approx \sqrt{2}\theta\left(E\pm \frac{\omega}{2}(1-\lambda)\right) u_L \komma \\ 
	u^\dagger_L \, \varepsilon(\lambda,\theta,\phi) \cdot \bar \sigma \, (p\pm k(\omega,\theta,\phi)) \cdot  \sigma \approx \sqrt{2}\theta\left(E\pm \frac{\omega}{2}(1+\lambda)\right) u^\dagger_L \label{LNidentity4} \komma
\end{gather}
holds. These identities (also see \cite{LN}) will be used in the amplitudes \eqref{amprawR1} to \eqref{amprawL4}. Interesting is that there is no $\phi$ or $\phis$ dependence in the expressions \eqref{LNidentity1} to \eqref{LNidentity4}. Furthermore, for a small arbitrary angle $\theta$ we have
\begin{gather}
	u^\dagger_R  \, \varepsilon(\lambda,\theta,\phi) \cdot \sigma \, u_R = u^\dagger_L  \, \varepsilon(\lambda,\theta,\phi) \cdot \bar \sigma \, u_L\approx  \sqrt{2} E \theta \punkt	
\end{gather}
The amplitudes from \eqref{amprawR1} to \eqref{amprawL4} simplify then to 
\begin{widetext}\begin{align}
	\im M_1^R &= -\im e^3 \frac{\theta\thetas\thetaq\left(2E+\omega(1+\lambda)\right)\left(2E+\omegas(1+\lambdas)\right)}{\sqrt{2}E\omega\omegas\left( \theta^2+\frac{m^2}{E^2} \right)\left( \thetas^2+\frac{m^2}{E^2} \right)} \label{ampapproxR1} \komma \\
	\im M_1^L &= -\im e^3 \frac{\theta\thetas\thetaq\left(2E+\omega(1-\lambda)\right)\left(2E+\omegas(1-\lambdas)\right)}{\sqrt{2}E\omega\omegas\left( \theta^2+\frac{m^2}{E^2} \right)\left( \thetas^2+\frac{m^2}{E^2} \right)} \label{ampapproxL1} \komma \\
	\im M_2^R &= -\im e^3 \frac{\theta\thetas\thetaq\left(2E-\omega(1-\lambda)\right)\left(2E-\omegas(1-\lambdas)\right)}{\sqrt{2}E\omega\omegas\left( \theta^2+\frac{m^2}{E^2} \right)\left( \thetas^2+\frac{m^2}{E^2} \right)} \label{ampapproxR2} \komma \\
	\im M_2^L &= -\im e^3 \frac{\theta\thetas\thetaq\left(2E-\omega(1+\lambda)\right)\left(2E-\omegas(1+\lambdas)\right)}{\sqrt{2}E\omega\omegas\left( \theta^2+\frac{m^2}{E^2} \right)\left( \thetas^2+\frac{m^2}{E^2} \right)} \label{ampapproxL2} \komma 
\end{align}
and
\begin{align}
	\im M_3^R &= \im e^3 \frac{\theta\thetas\thetaq\left(2E+\omega(1+\lambda)\right)\left(2E-\omegaq(1-\lambdaq)\right)}{\sqrt{2}E\omega\omegaq\left( \theta^2+\frac{m^2}{E^2} \right)\left( \thetaq^2+\frac{m^2}{E^2} \right)}  \label{ampapproxR3} \komma \\
	\im M_3^L &= \im e^3 \frac{\theta\thetas\thetaq\left(2E+\omega(1-\lambda)\right)\left(2E-\omegaq(1+\lambdaq)\right)}{\sqrt{2}E\omega\omegaq\left( \theta^2+\frac{m^2}{E^2} \right)\left( \thetaq^2+\frac{m^2}{E^2} \right)} \label{ampapproxL3} \komma \\
	\im M_4^R &= \im e^3 \frac{\theta\thetas\thetaq\left(2E-\omega(1-\lambda)\right)\left(2E+\omegaq(1+\lambdaq)\right)}{\sqrt{2}E\omega\omegaq\left( \theta^2+\frac{m^2}{E^2} \right)\left( \thetaq^2+\frac{m^2}{E^2} \right)} \label{ampapproxR4} \komma \\
	\im M_4^L &= \im e^3 \frac{\theta\thetas\thetaq\left(2E-\omega(1+\lambda)\right)\left(2E+\omegaq(1-\lambdaq)\right)}{\sqrt{2}E\omega\omegaq\left( \theta^2+\frac{m^2}{E^2} \right)\left( \thetaq^2+\frac{m^2}{E^2} \right)} \label{ampapproxL4} \punkt 
\end{align}\end{widetext}
Notice that the amplitudes of the left-handed electron just differ by exchanging all polarizations of the photons to minus the polarizations, i.e.
\begin{gather}
	\im M_i^L(\lambda,\lambdas,\lambdaq)=\im M_i^R(-\lambda,-\lambdas,-\lambdaq)\punkt\label{RLrelation}
\end{gather} 
Thus, apart from now we will write down only the amplitudes with the right-handed electron and get the amplitudes of the left-handed electron by this simple relation \eqref{RLrelation}.

We are interested in a very special corner of the phase space where $\thetas$ and $\thetaq$ are very small but still bigger than $\frac{m^2}{E^2}$, i.e.~$\thetas\gg\frac{m^2}{E^2}$ and $\thetaq\gg\frac{m^2}{E^2}$.\footnote{The reader may have noticed that in figure \ref{figure1} we omit two diagrams which are of the same topology. The reason is that these extra two diagrams have $1/(pk')(pq)$ propagators which won't lead to collinear divergence in this special corner of the phase space after applying the KLN theorem.} On the other side we allow $\theta$ to be of the order of $\frac{m^2}{E^2}$. Thus, $\thetas\gg\theta$ and $\thetaq\gg\theta$ holds as well. Furthermore, the phase space gets more restricted by the fact that the electron is forward scattered, i.e.~$p^\mu_\text{in}= p^\mu_\text{out}$. The constraint from energy and momentum conservation is then ${k'}^\mu= q^\mu+k^\mu$. This constraint in the collinear limit gives
\begin{gather}
	\omegas=\omegaq+\omega \text{ , } \thetas=\thetaq \frac{\omegaq}{\omega+\omegaq} \text{ and }\phis=\phiq\punkt
\end{gather}
The constraint $\phis=\phiq$ isn't important since these angles don't appear in the amplitudes \eqref{ampapproxR1} to \eqref{ampapproxL4}. Inserting the constraints and using the special corner of phase space one gets for the amplitudes
\begin{align}
	\im M_1^R &= -\im e^3 \frac{\theta\left(2E+\omega(1+\lambda)\right)\left(2E+(\omega+\omegaq)(1+\lambdas)\right)}{\sqrt{2}E\omega\omegaq\left( \theta^2+\frac{m^2}{E^2} \right)}\komma \\
	\im M_2^R &=  -\im e^3 \frac{\theta\left(2E-\omega(1-\lambda)\right)\left(2E-(\omega+\omegaq)(1-\lambdas)\right)}{\sqrt{2}E\omega\omegaq\left( \theta^2+\frac{m^2}{E^2} \right)}  \komma \\
	\im M_3^R &= \im e^3 \frac{\theta\left(2E+\omega(1+\lambda)\right)\left(2E-\omegaq(1-\lambdaq)\right)}{\sqrt{2}E\omega(\omega+\omegaq)\left( \theta^2+\frac{m^2}{E^2} \right)}   \komma \\
	\im M_4^R &= \im e^3 \frac{\theta\left(2E-\omega(1-\lambda)\right)\left(2E+\omegaq(1+\lambdaq)\right)}{\sqrt{2}E\omega(\omega+\omegaq)\left( \theta^2+\frac{m^2}{E^2} \right)}  \komma 
\end{align}
where we only kept the terms that will lead to collinear divergent terms after the phase space integration $\int\frac{\diff^3\vec k}{(2\pi)^3 	2\omega}$ of the incoming collinear photon. Then summing up the amplitudes gives
\begin{widetext}\begin{align}
	\im M^R = \sum_{i=1}^4\im M_i^R = - \im e^3\frac{\sqrt{2}\theta\left[ \omega(\omega+\omegaq)+(2E+\omega\lambda)(2E+(\omega+\omegaq)\lambdas) \right]}{E\omega\omegaq \left( \theta^2+\frac{m^2}{E^2} \right)} \nonumber  +\im e^3 \frac{\sqrt{2}\theta\left[-\omega\omegaq+(2E+\omega\lambda)(2E+\omegaq\lambdaq)\right]}{E\omega(\omega+\omegaq)\left(\theta^2+\frac{m^2}{E^2}\right)} \punkt  
\end{align}\end{widetext}
Clearly the collinear divergence comes from when one integrates the amplitude square over the $\theta$ since this is proportional to $\int_0^\delta\diff\theta\,\theta\frac{\theta^2}{\left(\theta^2+\frac{m^2}{E^2}\right)^2}\propto\logEdeltam$. We are interested in the full amplitude $\im M$, where we want to sum over the electron polarizations. In general holds for a generic amplitude $\bar u'^s \mathcal{M} u^r$ with an outgoing electron with spinor $\us$ and spin $s$ and an ingoing electron with spinor $u$ and spin $r$
\begin{widetext}\begin{gather}
	\frac{1}{4} \sum_{s,r=\pm\frac{1}{2}}\abs{\bar u'^s \mathcal{M} u^r}^2= \frac{1}{4}\sum_{s,r=\pm\frac{1}{2}} \bar u'^s \mathcal{M} u^r \bar u^r \mathcal{M}^\dagger u'^s =\nonumber \\
	\frac{1}{4}\left(\bar u'^{\frac{1}{2}} \mathcal{M} u^{\frac{1}{2}} \bar u^{\frac{1}{2}} \mathcal{M}^\dagger u'^{\frac{1}{2}}+\bar u'^{-\frac{1}{2}} \mathcal{M} u^{-\frac{1}{2}} \bar u^{-\frac{1}{2}} \mathcal{M}^\dagger u'^{-\frac{1}{2}}\nonumber \right. 
	\left.+\bar u'^{\frac{1}{2}} \mathcal{M} u^{-\frac{1}{2}} \bar u^{-\frac{1}{2}} \mathcal{M}^\dagger u'^{\frac{1}{2}} +\bar u'^{-\frac{1}{2}} \mathcal{M} u^{\frac{1}{2}} \bar u^{\frac{1}{2}} \mathcal{M}^\dagger u'^{-\frac{1}{2}}\right) \punkt \label{unpolarized}
\end{gather}\end{widetext}
The last line of equation \eqref{unpolarized} is the spin-flipping process of the amplitude or the helicity-flipping process in the collinear limit. Helicity-flipping processes don't possess collinear divergences, see e.g.~\cite{LN}. Thus interesting for us is the second line of \eqref{unpolarized}. Then, the unpolarized amplitude that will produce collinear divergences is given by
\begin{gather}\label{unpol}
	\frac{1}{4}\sum_\text{spins}\abs{\im M}^2 = \frac{1}{4} \left( \abs{\im M^R}^2 + \abs{\im M^L}^2 \right)\komma
\end{gather}
where $\im M^L = \sum_{i=1}^4\im M_i^L$.

Now we can apply the KLN-theorem and integrate over the phase space of the incoming photon $\int\frac{\diff^3\vec k}{(2\pi)^3 2\omega}$, which is in the collinear limit given by $\int \frac{\diff\omega\diff\theta\diff\phi\,\omega^2\sin\theta}{(2\pi)^32\omega}=\int\frac{\diff\omega\,\omega}{(2\pi)^2}\int_0^\delta\diff\theta\,\theta$, where we integrated $\int_0^{2\pi}\diff\phi=2\pi$ since the amplitudes do not depend on $\phi$. Then the collinear part of the unpolarized amplitude is 
\begin{widetext}\begin{gather}
	\int\frac{\diff^3\vec k}{(2\pi)^3 2\omega} \frac{1}{4}\sum_\text{spins}\abs{\im M}^2 =    \frac{1}{(2\pi)^3}\int \diff\omega \, \omega \int_0^\delta \diff\theta\,\theta \int_0^{2\pi} \diff\phi \, \frac{1}{4}\sum_\text{spins}\abs{\im M}^2  = \nonumber  \\
	\int\frac{\diff\omega\,\omega}{(2\pi)^2} \frac{e^6}{4E^2\omega^2}\logEdeltam \left[ \frac{-\omega\omegaq+(2E+\omega\lambda)(2E+\omegaq\lambdaq)}{(\omega+\omegaq)} -\frac{ \omega(\omega+\omegaq)+(2E+\omega\lambda)(2E+(\omega+\omegaq)\lambdas) }{\omegaq }   \right]^2 \nonumber \\
	+\left( \lambda\rightarrow-\lambda, \, \lambdas\rightarrow-\lambdas, \, \lambdaq\rightarrow-\lambdaq\right) \komma
\end{gather}\end{widetext}
which is the result we present in \eqref{ampsquare}.

\section{One-loop amplitude interfered with tree-level amplitude}\label{appendixC}
\begin{figure}[h]
	\centering
	\includegraphics[width=0.5 \textwidth]{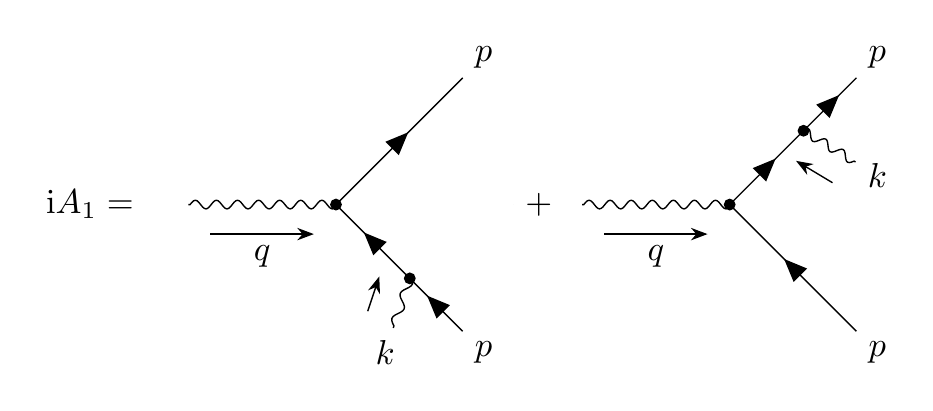}
	\caption{Tree-level diagrams with a collinear, incoming photon at order $e^2$.}	\label{fig1}
\end{figure}
\begin{figure}[h]
	\centering
	\includegraphics[width=0.5 \textwidth]{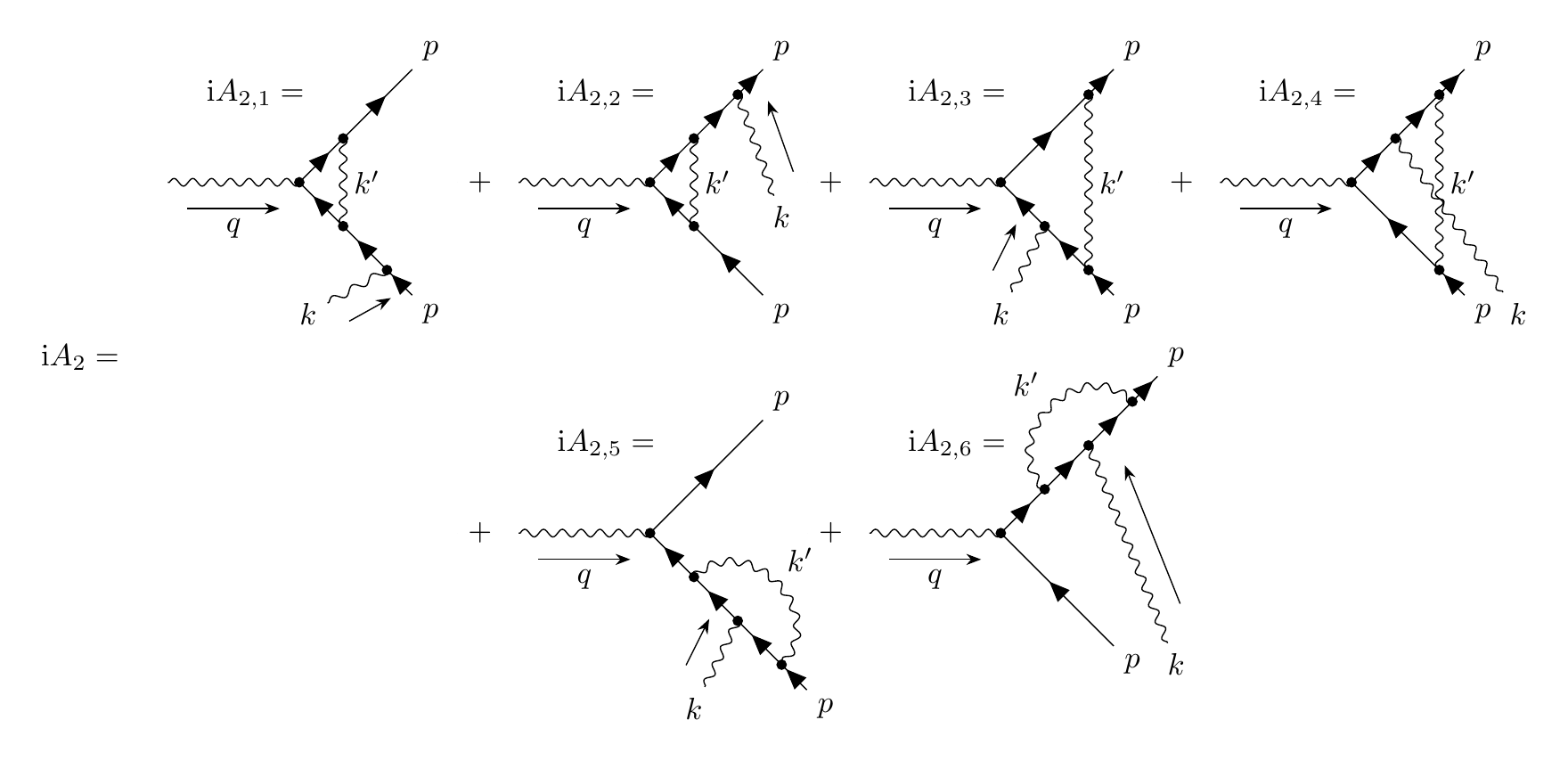}
	\caption{One-loop diagrams with a collinear, incoming photon at order $e^4$.} \label{fig2}
\end{figure}

An other term that contributes to the order $e^6$ in perturbation theory is the interference of the amplitudes in figure \ref{fig1} and \ref{fig2}. The process in figure \ref{fig1} describes an electron that scatters with two incoming photons, one with momentum $q^\mu$ which is the transfer momentum and of course $q^2=0$ holds as before and one with momentum $k^\mu$ which is a collinear photon. In order to possibly contribute to the cancelation process of the KLN anomaly in section \ref{KLNanomaly} the electron has to be forward scattered, thus $p^\mu_\text{in}=p^\mu_\text{out}$. Then from the conservation of energy and momentum we get the constraint $q^\mu=-k^\mu$, which means in the notation of appendix \ref{appendixB}: $\omegaq=-\omega$, $\thetaq=\theta$ and $\phiq=\phi$. The same constraint holds also for the amplitudes in the diagrams of figure \ref{fig2}. The amplitudes from figure \ref{fig2} are one-loop diagrams with a collinear incoming photon. We will change from now on the notation a little bit and name the amplitudes now $\im A$ instead of $\im M$ in order to keep it easier to distinguish but nevertheless keep the rest of the notations in appendix \ref{appendixB} the same. Of course when ever a photon runs in a loop it is no longer on-shell, i.e.~$k_\text{loop}^2\neq 0$. Other than in the appendix above we will write down the following amplitudes non-approximatively, i.e.~not in collinear limit, and just later when apply the KLN theorem we will Taylor expand the amplitudes in the collinear limit. 
\subsection{The tree-level amplitude at order $e^2$}
At this point we want to anticipate that the relation \eqref{RLrelation} holds here and in the following as well, which can't be seen directly but was found during the calculations for this appendix with the program Mathematica, i.e.
\begin{gather}
	\im A^L(\lambda,\lambda_q) = \im A^R(-\lambda,-\lambda_q) \label{RLsym} \punkt
\end{gather}
So we begin with the amplitude $\im A_1^R$ which is given by the diagrams in figure \ref{fig1} and gives
\begin{align}
	\im A_1^R  &= - \im e^2 \bar{u}_p^R \left[  \frac{\slashed{\varepsilon}_q (\slashed{p}+\slashed{k}) \slashed{\varepsilon}}{2pk} - \frac{\slashed{\varepsilon} (\slashed{p}-\slashed{k}) \slashed{\varepsilon}_q}{2pk} \right] u_p^R \nonumber \\ &= - \im e^2 u_R^\dagger \left[  \frac{\varepsilon_q \cdot \sigma \, (p+k)\cdot \bar \sigma  \, \varepsilon \cdot \sigma }{2pk} - \frac{\varepsilon\cdot\sigma\, (p-k) \cdot\bar\sigma \, \varepsilon_q\cdot \sigma}{2pk} \right] u_R \punkt \label{iA1R}
\end{align}
for the amplitude where a right-handed electron is forward scattered. 

Then the matrix multiplication in equation \eqref{iA1R} can be done by hand or using Mathematica and gives 
\begin{gather}\label{iA1}
	\im A_1^R = -\im e^2 \frac{4 E (1-\lambda \lambda_q\cos^2\theta	)\sin^2\left(\frac{\theta}{2}\right)}{(1+\cos^2\theta)(E-\abs{\vec{p}}\cos{\theta})}\punkt
\end{gather}
The amplitude for the left-handed electron is the same as the for the right-handed one, i.e.~$\im A_1^L=\im A_1^R$, since we have a multiplication of two polarizations $\lambda\lambdaq$. This is the first amplitude of the interference term that could cancel the KLN anomaly.

\subsection{One-loop amplitudes}\label{1loopamp}
The one-loop amplitudes of the diagrams in figure \ref{fig1} are given by
\begin{widetext}\begin{align}
	iA_{2,1}^{R/L} &= \frac{- e^4}{(2\pi)^4}\int \frac{\diff^4 k'}{k'^2}\frac{\bar u_p^{R/L} \, \gamma^\mu (\slashed p-\slashed k')\slashed \varepsilon_q (\slashed p -\slashed k'+\slashed k) \gamma_\mu (\slashed p+ \slashed k)\slashed\varepsilon \, u_p^{R/L}}{[(p-k'+k)^2-m^2][(p-k')^2-m^2][(p+k)^2-m^2]} \komma \label{iA21RL} \komma \\ 
	iA_{2,2}^{R/L} &= \frac{- e^4}{(2\pi)^4}\int \frac{\diff^4 k'}{k'^2}\frac{\bar u_p^{R/L} \,\slashed \varepsilon (\slashed p - \slashed k) \gamma^\mu (\slashed p-\slashed k'-\slashed k)\slashed \varepsilon_q (\slashed p -\slashed k') \gamma_\mu  \, u_p^{R/L}}{[(p-k'-k)^2-m^2][(p-k')^2-m^2][(p-k)^2-m^2]} \komma \\ 
	\im A_{2,3}^{R/L} &=  \frac{- e^4}{(2\pi)^4}\int \frac{\diff^4 k'}{k'^2}\frac{\bar u_p^{R/L} \, \gamma^\mu (\slashed p-\slashed k')\slashed \varepsilon_q (\slashed p -\slashed k'+\slashed k) \slashed \varepsilon(\slashed p- \slashed k') \gamma_\mu\, u_p^{R/L}}{[(p-k'+k)^2-m^2][(p-k')^2-m^2]^2}  \komma  \\ 
	\im A_{2,4}^{R/L} &= \frac{- e^4}{(2\pi)^4}\int \frac{\diff^4 k'}{k'^2}\frac{\bar u_p^{R/L} \, \gamma^\mu (\slashed p-\slashed k')\slashed \varepsilon (\slashed p -\slashed k'-\slashed k) \slashed \varepsilon_q(\slashed p- \slashed k') \gamma_\mu\, u_p^{R/L}}{[(p-k'-k)^2-m^2][(p-k')^2-m^2]^2}   \komma \\ 
	\im A_{2,5}^{R/L} &= \frac{- e^4}{(2\pi)^4}\int \frac{\diff^4 k'}{k'^2}\frac{\bar u_p^{R/L} \, \slashed\varepsilon_q (\slashed p+ \slashed k) \gamma^\mu (\slashed p -\slashed k'+\slashed k) \slashed \varepsilon (\slashed p-\slashed k')   \gamma_\mu  \, u_p^{R/L}}{[(p-k'+k)^2-m^2][(p-k')^2-m^2][(p+k)^2-m^2]}  \komma \\ 
	\im A_{2,6}^{R/L} &= \frac{- e^4}{(2\pi)^4}\int \frac{\diff^4 k'}{k'^2}\frac{\bar u_p^{R/L} \, \gamma^\mu (\slashed p -\slashed k') \slashed  \varepsilon (\slashed p-\slashed k'-\slashed k) \gamma_\mu (\slashed p - \slashed k) \slashed \varepsilon_q    \, u_p^{R/L}}{[(p-k'-k)^2-m^2][(p-k')^2-m^2][(p-k)^2-m^2]}\label{iA26RL} \punkt
\end{align}\end{widetext}

The denominator of the type $pk$ vanish in the collinear limit if $k^2$ is zero. So the interesting part of the one-loop amplitude is the one coming from the poles $pk'=0$ (collinearly) with $k'^2=0$. The pole $1/k'^2$ gives a contribution $\im \pi\delta(k'^2)$. As in \cite{Weinberg,YFS} shown the integral of $\int \diff \omegas^0$ in the amplitudes \eqref{iA21RL} to \eqref{iA26RL} sets the loop-photon with momentum $k'$ on-shell. We use the standard $\gamma$-matrices identity $\gamma^\mu \gamma^\alpha \gamma^\beta \gamma^\nu \gamma_\mu = -2\gamma^\nu \gamma^\beta \gamma^\alpha$ and for the amplitudes $\im A_{2,3}$ and $\im A_{2,4}$ we use a formula that can be easily can verified and only holds for the specific choice of spinors \eqref{spinors} and using $\gamma^\mu$ in Weyl representation: $\bar u_p^{R/L} \, \gamma^\mu [...] \gamma_\mu \, u_p^{R/L} = -2 \bar u_p^{L/R}  [...]  u_p^{L/R}$, where $[...]$ stands for any set of $\gamma$-matrices.

Then the amplitudes \eqref{iA21RL} to \eqref{iA26RL} are given by
\begin{widetext}\begin{gather}
	\im A_{2,1}^{R/L} = \frac{- \im e^4}{(2\pi)^3}\int \frac{\diff^3 k'}{\omega'}\frac{ \bar u_p^{R/L} \, (\slashed p -\slashed k'+\slashed k) \slashed \varepsilon_q (\slashed p-\slashed k') (\slashed p+ \slashed k)\varepsilon \, u_p^{R/L}}{[-2pk'+2pk-2kk'](2pk')(2pk)} \label{iA21} \komma \\
	\im A_{2,2}^{R/L} = \frac{ \im e^4}{(2\pi)^3}\int \frac{\diff^3 k'}{\omega'}\frac{\bar u_p^{R/L} \,\slashed \varepsilon (\slashed p - \slashed k)  (\slashed p -\slashed k') \slashed \varepsilon_q (\slashed p-\slashed k'-\slashed k)  \, u_p^{R/L}}{[-2pk'-2pk+2kk'](2pk')(2pk)} \komma \\ 
	\im A_{2,3}^{R/L} = \frac{\im e^4}{(2\pi)^3}\int \frac{\diff^3 k'}{\omega'}\frac{\bar u_p^{L/R} \, (\slashed p-\slashed k')\slashed \varepsilon_q (\slashed p -\slashed k'+\slashed k) \slashed \varepsilon(\slashed p- \slashed k') \, u_p^{L/R}}{[-2pk'+2pk-2kk'](2pk')^2}  \label{iA23}  \komma \\ 
	\im A_{2,4}^{R/L} = \frac{\im e^4}{(2\pi)^3}\int \frac{\diff^3 k'}{\omega'}\frac{ \bar u_p^{L/R} \,  (\slashed p-\slashed k')\slashed \varepsilon (\slashed p -\slashed k'-\slashed k) \slashed \varepsilon_q(\slashed p- \slashed k') \, u_p^{L/R}}{[-2pk'-2pk+2kk'](2pk')^2}  \label{iA24}  \komma \\ 
	\im A_{2,5}^{R/L} = \frac{- \im e^4}{(2\pi)^3}\int \frac{\diff^3 k'}{\omega'}\frac{\bar u_p^{R/L} \, \slashed\varepsilon_q (\slashed p+ \slashed k)  (\slashed p-\slashed k') \slashed \varepsilon (\slashed p -\slashed k'+\slashed k)  \, u_p^{R/L}}{[-2pk'+2pk-2kk'](2pk')(2pk)}  \komma \\ 
	\im A_{2,6}^{R/L} = \frac{ \im e^4}{(2\pi)^3}\int \frac{\diff^3 k'}{\omega'}\frac{\bar u_p^{R/L} \,  (\slashed p-\slashed k'-\slashed k) \slashed  \varepsilon (\slashed p -\slashed k')  (\slashed p - \slashed k) \slashed \varepsilon_q  \, u_p^{R/L}}{[-2pk'-2pk+2kk'](2pk')(2pk)} \label{iA26}\punkt
\end{gather}\end{widetext}

\paragraph*{Comment on the amplitudes $\im A_{2,3}^R$ and $ \im A_{2,4}^R$:}
The formulas \eqref{iA23} and \eqref{iA24} for the amplitudes $\im A_{2,3}^R$ and $ \im A_{2,4}^R$ show that the incoming photon with momentum $k^\mu$ is non-IR absorption in these two amplitudes, since this photon is attached to an internal line so that there is no propagator with $1/pk$ in the amplitudes (see for example \cite{Weinberg,YFS,us2}).

\subsubsection{Symmetries between the amplitudes $\im A_{2,i}^{R/L}$}
If one takes a closer look to the amplitudes \eqref{iA21} to \eqref{iA26} one can manifest some symmetries between them. $\im A_{2,3}$ and $\im A_{2,4}$ are related to each other as well as the other 4 amplitudes. In the following we will omit the right-/left-labelling of the amplitudes to keep it shorter. Writing the amplitudes as functions of the polarisations of the photons $\lambda$, $\lambda_q$ and the energy of the incoming collinear photon $\omega$ then 
\begin{widetext}\begin{align}
	\im A_{2,4}(\lambda,\lambda_q,\omega) &= \im A_{2,3}(\lambda_q,\lambda,-\omega)\komma \label{iA23sym1} \\ 
	\im A_{2,4}(\lambda,\lambda_q,\omega) &= \im A^\dagger_{2,3}(-\lambda,-\lambda_q,-\omega) =  \im A_{2,3}(-\lambda,-\lambda_q,-\omega)\komma \label{iA23sym2}
\end{align}\end{widetext}
holds for $\im A_{2,4}$. For the other amplitudes one can show that
\begin{widetext}\begin{align}
	\im A_{2,2} (\lambda,\lambda_q,\omega)&= \im A^\dagger_{2,1}(-\lambda,-\lambda_q,-\omega) = \im A_{2,1}(-\lambda,-\lambda_q,-\omega) \komma \\ 
	\im A_{2,5} (\lambda,\lambda_q,\omega)&= \im A_{2,2}(\lambda_q,\lambda,-\omega) = \im A_{2,1}(-\lambda_q,-\lambda,\omega) \komma \\ 
	\im A_{2,6} (\lambda,\lambda_q,\omega)&= \im A_{2,1}(\lambda_q,\lambda,-\omega) \komma
\end{align}\end{widetext}
holds. We want to anticipate that the integrals in \eqref{iA21} to \eqref{iA26} are real, which can't be seen directly since the numerators have terms proportional to $\e{\pm\im n (\phi'-\phi)}$, with $n=0,1,2$, but this showed up during the calculations for this appendix.

Together with relation \eqref{RLsym} one only has to calculate $\im A_{2,1}^R$ and $\im A_{2,3}^R$ to get the complete amplitude $\im A_2^{R/L} = \sum_{i=1}^6 \im A_{2,i}^{R/L}$. There are two other symmetry in this amplitude. The first is changing $\lambda\rightarrow-\lambda$, $\lambda_q\rightarrow-\lambda_q$ and $\omega\rightarrow-\omega$, which is nothing else but having outgoing photons in figure \ref{fig2}. The second is changing $\lambda\rightarrow-\lambda_q$ and $\lambda_q\rightarrow-\lambda$, which is exchanging the ingoing photon with momentum $k^\mu$/$q^\mu$ to an outgoing photon with momentum $q^\mu$/$k^\mu$. In other words, the amplitudes of the diagrams with outgoing photons instead of ingoing once are the same as the amplitudes in figure \ref{fig2}. This is the same behaviour as already seen for the IR case in \cite{Weinberg,LN,mcmullan}. The symmetries can be seen in
\begin{widetext}\begin{gather}
	\im A_2(\lambda,\lambda_q,\omega) = \sum_{i=1}^6 \im A_{2,i}(\lambda,\lambda_q,\omega) \nonumber \\	
	 = \im A_{2,1}(\lambda,\lambda_q,\omega)+\im A_{2,1}(-\lambda,-\lambda_q,-\omega)+\im A_{2,1}(-\lambda_q,-\lambda,\omega)+\im A_{2,1}(\lambda_q,\lambda,-\omega)\nonumber  \\ 
	 +\im A_{2,3}(\lambda,\lambda_q,\omega)+\im A_{2,3}(\lambda_q,\lambda,-\omega)  \\ 
	= \im A_{2,1}(\lambda,\lambda_q,\omega)+\im A_{2,1}(-\lambda,-\lambda_q,-\omega)+\im A_{2,1}(-\lambda_q,-\lambda,\omega)+\im A_{2,1}(\lambda_q,\lambda,-\omega)\nonumber \\
	 +\im A_{2,3}(\lambda,\lambda_q,\omega)+\im A_{2,3}(-\lambda,-\lambda_q,-\omega) \komma
\end{gather}\end{widetext}
where the last two lines are related by \eqref{iA23sym1} and \eqref{iA23sym2}.

The two determining amplitudes are \eqref{iA21} and \eqref{iA23} which can be simplified to
\begin{widetext}\begin{gather}
	\im A_{2,1}^{R} =  \frac{- \im e^4}{(2\pi)^3}\int \frac{\diff^3 k'}{\omega'}\frac{  u_{R}^\dagger \, ( p - k'+ k)\cdot \sigma  \varepsilon_q \cdot\bar\sigma ( p- k')\cdot\sigma ( p+  k)\cdot\bar\sigma\varepsilon\cdot \sigma\, u_{R}}{[-2pk'+2pk-2kk'](2pk')(2pk)} \komma \\ 
	\im A_{2,3}^{R} =  \frac{ \im e^4}{(2\pi)^3}\int \frac{\diff^3 k'}{\omega'}\frac{  u_{L}^\dagger \, ( p - k')\cdot \bar \sigma  \varepsilon_q \cdot\sigma ( p- k'+k)\cdot\bar\sigma \varepsilon\cdot \sigma ( p-  k')\cdot\bar\sigma\, u_{L}}{[-2pk'+2pk-2kk'](2pk')^2} \punkt
\end{gather}\end{widetext}
We can now move forward to perform the integral $\int\frac{\diff^3k'}{\omega'} = \int\diff\omega'\diff\theta'\diff\phi' \omega'\sin\theta'$, where we will see that in fact the amplitudes $A_{2,i}$ are real after the integration.

\subsubsection{Details of the $\int_0^{2\pi}\diff \phi'$ integral}
In the denominator $\phi'$ appears only in the $kk'=k\cdot k'$, where $k$ and $k'$ are on-shell. The nominators of $\im A_{2,1}^R$ and $\im A_{2,3}^R$ are calculated by Mathematica and have terms that go like $\e{\pm \im n (\phi'-\phi)}$, with $n=0,1,2$. Then there are integrals of the form
\begin{gather}
	\int_0^{2\pi} \frac{\e{\pm\im n (\phi'-\phi)}}{a+b\cos(\phi'-\phi)}\diff \phi' \komma
\end{gather}
where $a$ and $b$ are independent of $\phi'$ and $\phi$. All integrals including $\sin(n (\phi'-\phi))$ vanish, as well as the one with $n=0$, i.e.~the one with a constant term. The non-vanishing integrals are
\begin{gather}
	\int_0^{2\pi} \diff \phi' \frac{\cos( \phi'-\phi)}{a+b\cos(\phi'-\phi)} = \frac{2\pi}{b}\komma \\ 
	\int_0^{2\pi} \diff \phi' \frac{\cos( 2(\phi'-\phi))}{a+b\cos(\phi'-\phi)} = -\frac{4\pi a}{b^2} \punkt
\end{gather}
In the integration the variables are $a=-2pk'+2pk-\omega\omega'(1-\cos\theta \cos\theta')$ and $b=2\omega \omega' \sin\theta\sin\theta'$. We won't write down the amplitudes $\im A_{2,1}^R$ and $\im A_{2,3}^R$ after the $\int_0^{2\pi}\diff\phi'$ integration, since these terms are quite long and there is no greater benefit from knowing these formulas. So that we go on to the next integration.

\subsubsection{Details of the $\int\diff \theta' $ integral}\label{thetasintegral}
The next step is the $\int\diff\theta'$ integral, where the $\log(m_e)$ divergences will appear. In the following we will keep terms that are finite after the $\int\diff\theta'$ integration and the terms that are logarithmic divergent. What we will omit are terms that go as the mass of the electron $m_e$ since they will vanish in the collinear limit where $m_e\rightarrow 0$. In the logarithmic divergent terms we will have to use an angle regulator $\delta$ which is a small angle between the electron and the collinear photon in the loop.\footnote{As it was the case in appendix \ref{appendixB}, where we also had to put a regulator for the angle between the electron and the absorbed collinear photon connected to the external line of the electron.} In the finite terms we can perform the full integration, meaning integrate over $\theta'$ from $0$ to $\pi$. In the following we will skip some intermediate steps and write down the amplitude $\im A_2^R$ after the two angle integrations $\int\diff\thetas\diff\phi'\sin\thetas$, since the formula for the full amplitude $\im A_2^R$ is shorter than all the single amplitudes $\im A_{2,i}^R$. Then the integrals are performed by Mathematica and give the result  
\begin{widetext}\begin{gather}
	\im A_2^R =\frac{\im e^4}{(2\pi)^3} \int \diff\omega'  \frac{4\pi}{\omega^2(E-\abs{\vec p}\cos\theta)(1+\cos^2\theta)(1+\cos\theta)}	\left\{ \omega' \cos\theta \left[ 2E(1+(2+\lambda\lambda_q)\cos\theta)-\omega(\lambda-\lambda_q)(1-\cos^2\theta) \right]   \right. \nonumber \\
	\left.  +\omega \sin^2\theta \left[ E(\lambda-\lambda_q)\cos\theta + 2\omega (1-\lambda\lambda_q\cos^2\theta) \right]\log\left(\frac{E\delta}{m}\right) \right\} + \nonumber \\ 
	+\frac{\im e^4}{(2\pi)^3} \int \diff\omega' \frac{2\pi}{E\omega(1+\cos^2\theta)}\left\{ \left[ E(\lambda-\lambda_q)\cos\theta + \omega(1-\lambda\lambda_q\cos^2\theta)\right] \left( 1-2\logEdeltam \right)  + \frac{4E\omega'}{\omega}(1+\lambda\lambda_q)\cot^2\theta \right\} \label{iA2R} \punkt
\end{gather}\end{widetext}
The first three lines come from $\im A_{2,1}^R+\im A_{2,2}^R+\im A_{2,5}^R+\im A_{2,6}^R$, which can be seen from the factor $pk=\omega(E-\abs{\vec p}\cos\theta)$ in the denominator, and the fourth and the fifth line come from $\im A_{2,3}^R + \im A_{2,4}^R$. What also can be seen is that there is no IR divergence in the one-loop amplitude after performing the $\int\diff\omega'$ integration, which is conform with \cite{YFS,Weinberg} since the $B$-factor vanishes in the forward scattering. 

\subsection{The interference term}\label{interference}
In order to see if there is a cancelation with \eqref{ampsquare} to the order $e^6$ one has to apply the KLN theorem to the unpolarized interference term of the amplitude $\im A_1$ and $\im A_2$, i.e.~$\im A_1^R \left(\im A_2^R\right)^*+\im A_1^L \left(\im A_2^L\right)^*+\text{h.c.}$, where we used \eqref{unpol}. The contribution is given by 
\begin{gather}\label{unpolIntterm}
	\frac{1}{4} \int \frac{\diff^3\vec k}{(2\pi)^3 2\omega}\left[\im A_1^R \left(\im A_2^R\right)^*+\im A_1^L \left(\im A_2^L\right)^*+\text{h.c.} \right]\punkt
\end{gather}
In the following we again just calculate the contribution coming from the $\im A_1^R \left(\im A_2^R\right)^*$ since we can apply the relation \eqref{RLsym} to get the contribution coming from the amplitude with the left-handed electron.

\subsubsection{Small angle approximation}\label{smallangle}
A Taylor expansion for small $\theta$ of the expressions in \eqref{iA1} and \eqref{iA2R} gives
\begin{widetext}\begin{align}
	\im A_1^{R} \approx& -\im e^2 (1-\lambda\lambda_q) \frac{\theta^2 }{\theta^2+\frac{m^2}{E^2}}\komma \label{iA1approx}\\
	\im A_2^{R} \approx& \, \frac{\im e^4}{(2\pi)^2} \int \diff \omega'  \frac{1}{\omega^2} \left\{ 2 \omega' \, \frac{3+\lambda\lambda_q}{\theta^2+\frac{m^2}{E^2}} +\frac{\theta^2}{\theta^2+\frac{m^2}{E^2}} \left[-\frac{\omega'}{2E}\left(2\omega(\lambda-\lambda_q)+E(1+\lambda\lambda_q)\right)\nonumber \right.\right. \\ 
	&\left.\left.+\frac{\omega}{E}\left(2\omega(1-\lambda\lambda_q)+E(\lambda-\lambda_q)\right)\logEdeltam \nonumber \right]\right\}  \\ 
	&+\frac{\im e^4}{(2\pi)^2} \int \diff\omega'  \left[ \frac{\lambda-\lambda_q}{\omega} +\frac{1-\lambda\lambda_q}{E}\right] \left( \frac{1}{2}-\logEdeltam \right) +\frac{\theta^2(1+\lambda\lambda_q)}{4}\left[ \frac{1-2\logEdeltam}{E}+\frac{22}{15}\frac{\omega'}{\omega^2} \right]  \nonumber \\ 
	&\left. +\frac{\omega'(1+\lambda\lambda_q)}{3\omega^2}  - \frac{2\omega'(1+\lambda\lambda_q)}{\omega^2\theta^2}    \right\} \label{iA2approx} \punkt
\end{align}\end{widetext}
where as usual in the collinear limit (see \cite{LN}) $2pk = 2\omega (E-\abs{\vec p}\cos\theta) \approx \omega E (\theta^2+m^2/E^2)$. 
 
From the Taylor expansion of the two amplitudes we can see that once the KLN theorem is applied the interference term will have collinear divergences coming from the term proportional to $\theta^2/(\theta^2+m^2/E^2)^2$. And there will be terms that are collinearly divergent coming only from the loop integration from the previous section \ref{thetasintegral}.

The interference term of $\im A_1^R$ and $\left(\im A_2^R\right)^*$ involves the following multiplication of terms with the polarisations of the photons
\begin{gather}
	(1-\lambda\lambda_q)(3+\lambda\lambda_q) = 2(1-\lambda\lambda_q) \komma \\ 
	(1-\lambda\lambda_q)(1+\lambda\lambda_q) = 0 \komma \\
	(1-\lambda\lambda_q)(\lambda-\lambda_q) = 2(\lambda-\lambda_q) \komma \\ 
	(1-\lambda\lambda_q)(1-\lambda\lambda_q) = 2(1-\lambda\lambda_q) \komma
\end{gather}
where $\lambda^2=1$ and $\lambda_q^2=1$ is used. The integral of interest is then
\begin{widetext}\begin{gather}
	\int \frac{\diff^3\vec k }{(2\pi)^3 2\omega} \left(\im A_1^R \left(\im A_2^R\right)^*+\text{h.c.}\right)=\int \diff\omega\,\omega\int\limits_0^{\delta}\diff\theta\,\theta\int\limits_0^{2\pi}\diff\phi \left(\im A_1^R \left(\im A_2^R\right)^*+\text{h.c.}\right) \approx \nonumber\\
	   - \frac{ e^6}{(2\pi)^4}  \int\diff\omega  \int\diff \omega' \int\limits_0^\delta\diff\theta  \frac{\theta}{\omega} \left\{ 2 \omega' \, \frac{2(1-\lambda\lambda_q)\theta^2}{\left(\theta^2+\frac{m^2}{E^2}\right)^2} +\frac{\theta^4}{\left(\theta^2+\frac{m^2}{E^2}\right)^2} \left[-2\frac{\omega\omega'}{E}(\lambda-\lambda_q)\nonumber +2\frac{\omega}{E}\left(2\omega(1-\lambda\lambda_q)+E(\lambda-\lambda_q)\right)\logEdeltam \nonumber \right]\right\} \\ 
	+\frac{ e^6}{(2\pi)^4}\int\diff\omega \int \diff\omega'  \int\limits_0^\delta\diff\theta\,\theta  \frac{\theta^2 }{\theta^2+\frac{m^2}{E^2}} \left[ \lambda-\lambda_q +\frac{\omega}{E}(1-\lambda\lambda_q)\right] \left( 1-2\logEdeltam \right)  \punkt
\end{gather}\end{widetext}
Then one can use the following identities 
\begin{gather}	\int_0^\delta \diff\theta \, \theta  \frac{\theta^2}{(\theta^2+\frac{m^2}{E^2})^2} = -\frac{1}{2}+\logEdeltam \komma \end{gather}
\begin{gather} \int_0^\delta \diff\theta \, \theta  \frac{\theta^4}{(\theta^2+\frac{m^2}{E^2})^2} = \frac{\delta^2}{2} \komma\end{gather}
\begin{gather} \int_0^\delta \diff\theta \, \theta  \frac{\theta^2}{\theta^2+\frac{m^2}{E^2}} = \frac{\delta^2}{2} \komma  \end{gather}
to simplify the interference term
\begin{widetext}\begin{gather}
	\int \frac{\diff^3\vec k }{(2\pi)^3 2\omega} \left(\im A_1^R \left(\im A_2^R\right)^*+\text{h.c.}\right)\approx	\nonumber\\
	   \frac{ e^6}{(2\pi)^4}  \int\diff\omega  \int\diff \omega' \,  \left\{ - 2 \frac{\omega'}{\omega} \, (1-\lambda\lambda_q)\left(1-2\logEdeltam\right) +\delta^2  \left[-\frac{\omega'}{E}(\lambda-\lambda_q)\nonumber +\frac{1}{E}\left(2\omega(1-\lambda\lambda_q)+E(\lambda-\lambda_q)\right)\logEdeltam \nonumber \right]\right\} \\ 
	+ \frac{ e^6}{(2\pi)^4}\int\diff\omega \int \diff\omega'  \,  \delta^2 \left[ \lambda-\lambda_q +\frac{\omega}{E}(1-\lambda\lambda_q)\right] \left( \frac{1}{2}-\logEdeltam \right)   \komma
\end{gather}
which is more simplified 
\begin{gather}
	\int_0^\delta \frac{\diff^3 \vec k }{(2\pi)^3 2\omega} \left(\im A_1^R \left(\im A_2^R\right)^*+\text{h.c.}\right)	\approx \nonumber\\
	  -  \frac{ e^6}{(2\pi)^4}  \int\diff\omega  \int\diff \omega' \, \left\{ 2\frac{\omega'}{\omega}(1-\lambda\lambda_q)\left( 1-2\logEdeltam \right)  - \frac{\delta^2}{2}\left[ \frac{\omega}{E}(1-\lambda\lambda_q)\left(1+2\logEdeltam\right)+\left(1-2\frac{\omega'}{E}\right)(\lambda-\lambda_q) \right]\right\}\label{int-term-approx} \punkt
\end{gather}\end{widetext}

The $\log$ term in the second line of \eqref{int-term-approx} is coming from the phase space integration $\int\diff^3\vec k/\omega$. There are no $\log^2$ terms and the IR divergent term is coming from the $\int\diff \omega/\omega$ integration not from the loop integral $\int\diff\omega'$. This is conform with  \cite{us2,Weinberg,YFS}, since the IR term comes from the interference term of the amplitudes $\im A_1$ and $\im A_{2,1}+\im A_{2,2}+\im A_{2,5}+\im A_{2,6}$, which are the amplitudes where the incoming photon is attached to a external electron line. 

\subsubsection{Full interference term}
The unpolarized contribution is given by \eqref{unpolIntterm} which is now
\begin{widetext}\begin{gather}
	\frac{1}{4}  \int \frac{\diff^3\vec k}{(2\pi)^3 2\omega} \left[\im A_1^R(\lambda,\lambda_q) \left(\im A_2^R(\lambda,\lambda_q)\right)^*+\im A_1^R(-\lambda,-\lambda_q) \left(\im A_2^R(-\lambda,-\lambda_q)\right)^*+\text{h.c.}\right] = \nonumber \\
	-  \frac{ e^6}{(2\pi)^4}  \int\diff\omega  \int\diff \omega' \, \left\{ 4\frac{\omega'}{\omega}(1-\lambda\lambda_q)\left( 1-2\logEdeltam \right) - \delta^2\left[ \frac{\omega}{E}(1-\lambda\lambda_q)\left(1-2\logEdeltam\right) \right]\right\} \punkt
\end{gather}\end{widetext}
The collinear divergent part is given by 
\begin{gather}
	  \frac{ e^6}{8\pi^4}  \int\diff\omega  \int\diff \omega' \, \left\{ 4\frac{\omega'}{\omega} -\delta^2 \frac{\omega}{E} \right\} (1-\lambda\lambdaq)\logEdeltam \punkt	  
\end{gather}
This term does not cancel the KLN anomaly in equation \eqref{ampsquare}.

\section{Two-loop amplitude}\label{appendixD}
\begin{figure}[h]
	\centering
	\includegraphics[width=0.5 \textwidth]{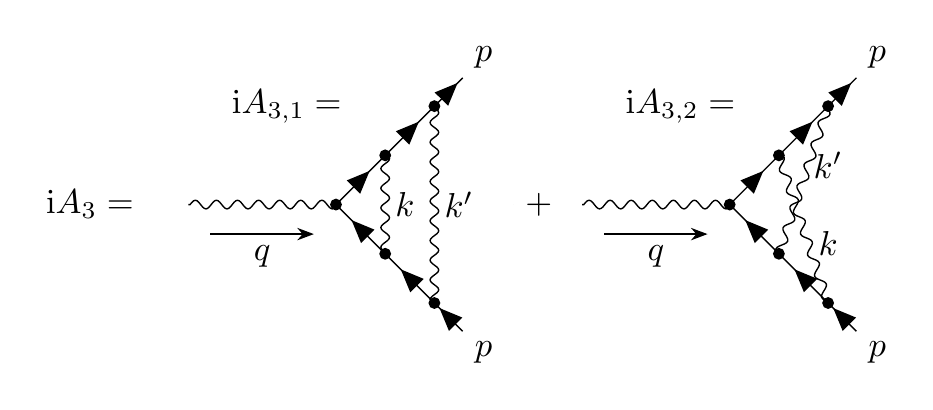}
	\caption{Two-loop diagrams at order $e^5$ for a forward scattered electron with 4-momentum $p^\mu$.} \label{fig4}
\end{figure}

There are also possible cancelation with two-loop diagrams, since an interference term of an amplitude of order $e$ and the two-loop amplitude in figure \ref{fig4} (order $e^5$) is again of order $e^6$. 
\newline
\subsection{Amplitude $\im A_{3,1}$}
From the diagram in figure \ref{fig4} one can read off
\begin{widetext}\begin{gather}\label{iA41old}
	\im A_{3,1}^{R/L} = \frac{\im e^5}{(2\pi)^8} \int\frac{\diff^4k\,\diff^4k'}{k^2\, k'^2} \frac{\bar u_p^{R/L}\gamma^\mu(\slashed{p}-\slashed{k}')\gamma^\nu(\slashed{p}-\slashed{k}'-\slashed{k})\varepsilon_q(\slashed{p}-\slashed{k}'-\slashed{k})\gamma_\nu (\slashed{p}-\slashed{k}')\gamma_\mu u_p^{R/L}}{\left[(p-k'-k)^2-m^2\right]^2\left[(p-k')^2-m^2\right]^2} \punkt
\end{gather}\end{widetext}
To simplify the amplitude we use the same steps and identities of section \ref{1loopamp}. Then the amplitude $\im A_{4,1}^{R/L}$ is simplified to 
\begin{widetext}\begin{gather} \label{iA41}
	\im A_{4,1}^{R/L} = \frac{-\im e^5}{(2\pi)^6} \int\frac{\diff^3k\,\diff^3k'}{\omega  \omega'} \frac{\bar u_p^{L/R}(\slashed{p}-\slashed{k}')(\slashed{p}-\slashed{k}'-\slashed{k})\slashed \varepsilon_q(\slashed{p}-\slashed{k}'-\slashed{k})(\slashed{p}-\slashed{k}')u_p^{L/R}}{\left[-2pk'-2pk+2kk'\right]^2\left(2pk'\right)^2} \punkt
\end{gather}\end{widetext}
As before we will look first at the $\int_0^{2\pi}\diff\phi'$ integration. The integration is a bit different since the denominator is of the form $\left[a+b\cos(\phi'-\phi)\right]^2$, where $a$ and $b$ are again independent of $\phi$ and $\phi'$. Mathematica calculates the nominator inside the integrals of the amplitude \eqref{iA41} and what we get are terms proportional to the exponential functions $\e{\pm \im(\phi-\phi_q)}$, $\e{\pm\im(\phi'-\phi_q)}$, $\e{\pm\im(\phi'-\phi)}$ and $\e{\pm\im(2\phi-\phi'-\phi_q)}$.\footnote{Notice that the angles of $\theta$ and $\phi$ are the ones of the on-shell photon that runs in the loop with momentum $k$.} The amplitude $\im A_{3,1}^{R/L}$ vanishes, since 
\begin{gather} \int_0^{2\pi} \diff\phi' \, \frac{\e{\pm \im(\phi-\phi_q)}}{\left[a+b\cos(\phi'-\phi)\right]^2} = 0 \komma \end{gather} 
\begin{gather}\int_0^{2\pi} \diff\phi' \, \frac{\e{\pm \im(\phi'-\phi_q)}}{\left[a+b\cos(\phi'-\phi)\right]^2} = 0 \komma \end{gather} 
\begin{gather}\int_0^{2\pi} \diff\phi' \, \frac{\e{\pm \im(\phi'-\phi)}}{\left[a+b\cos(\phi'-\phi)\right]^2} = 0 \komma \end{gather} 
\begin{gather}\int_0^{2\pi} \diff\phi' \, \frac{\e{\pm\im(2\phi-\phi'-\phi_q)}}{\left[a+b\cos(\phi'-\phi)\right]^2} = 0 \punkt \end{gather}
Thus, there can only be a contribution from the other diagram in figure \ref{fig4}.
\subsection{Amplitude $\im A_{3,2}$}
From the figure \ref{fig4} one can read off that the amplitude is given by
\begin{widetext}\begin{gather}\label{iA42old}
	\im A_{3,2}^{R/L} =  \frac{\im e^5}{(2\pi)^8} \int\frac{\diff^4k\,\diff^4k'}{k^2\, k'^2} \frac{\bar u_p^{R/L}\gamma^\mu(\slashed{p}-\slashed{k}')\gamma^\nu(\slashed{p}-\slashed{k}'-\slashed{k})\slashed\varepsilon_q(\slashed{p}-\slashed{k}'-\slashed{k})\gamma_\mu (\slashed{p}-\slashed{k})\gamma_\nu u_p^{R/L}}{\left[(p-k'-k)^2-m^2\right]^2\left[(p-k')^2-m^2\right]\left[(p-k)^2-m^2\right]} \punkt
\end{gather}\end{widetext}
Notice the difference to \eqref{iA41old}. The two last $\gamma$-matrices are exchanged and in the denominator there is a propagator $pk$. This differences makes it impossible to directly apply the identities of section \ref{1loopamp}. Before using them we will apply an other $\gamma$-matrix identity, $\gamma^\mu\gamma^\nu\gamma^\rho = \eta^{\mu\nu}\gamma^\rho+ \eta^{\nu\rho}\gamma^\mu- \eta^{\mu\rho}\gamma^\nu-\im\varepsilon^{\sigma\mu\nu\rho}\gamma_\sigma\gamma^5$, where $\eta^{\mu\nu}$ is the metric tensor in Minkowski spacetime, $\varepsilon^{\sigma\mu\nu\rho}$ is the Levi-Civita symbol in 4d. With that we rewrite the middle part of the nominator to
\begin{widetext}\begin{gather}\label{id3}
	(\slashed{p}-\slashed{k}'-\slashed{k})\varepsilon_q(\slashed{p}-\slashed{k}'-\slashed{k}) = 2(p-k'-k)\cdot \varepsilon_q \, (\slashed{p}-\slashed{k}'-\slashed{k}) - (p-k'-k)^2 \,\slashed\varepsilon_q \punkt
\end{gather}\end{widetext}
The term with the Levi-Civita symbol is zero since there is a summation of 2 equal terms, i.e.~$\varepsilon^{\alpha\beta\mu\nu}a_\alpha a_\beta b_\mu c_\nu= 0$. Once this identity is used one can also apply the identities of section \ref{1loopamp}. Putting all together then the amplitude \eqref{iA42old} is
\begin{widetext}\begin{align}
	\im A_{3,2}^{R/L} = &-\frac{\im e^5}{(2\pi)^6} \int\frac{\diff^3k\,\diff^3k'}{2\omega\, \omega'} (p-k'-k)\cdot \varepsilon_q  \frac{\bar u_p^{R/L}\gamma^\mu  (\slashed p-\slashed k')\gamma^\nu (\slashed{p}-\slashed{k}'-\slashed{k}) \gamma_\mu (\slashed{p}-\slashed{k})\gamma_\nu u_p^{R/L}}{\left[-2pk'-2pk+2kk'\right]^2(2pk')(2pk)} \nonumber \\
	&+\frac{\im e^5}{(2\pi)^6} \int\frac{\diff^3k\,\diff^3k'}{2\omega\, 2\omega'}  \frac{\bar u_p^{R/L}\gamma^\mu (\slashed p-\slashed k')\gamma^\nu \slashed\varepsilon_q \gamma_\mu (\slashed{p}-\slashed{k})\gamma_\nu u_p^{R/L}}{\left[-2pk'-2pk+2kk'\right](2pk')(2pk)} \komma
\end{align}\end{widetext}
where in the second line one $-2pk'-2pk+2kk'$ propagator is canceled by the $(p-k'-k)^2=-2pk'-2pk+2kk'$ term in the identity \eqref{id3}. Now we use the identities $\gamma^\mu \gamma^\alpha \gamma^\beta \gamma^\nu \gamma_\mu = -2\gamma^\nu \gamma^\beta \gamma^\alpha$ and $\gamma^\mu \gamma^\alpha \gamma^\beta \gamma_\mu = 4\eta^{\alpha\beta}$ to get
\begin{widetext}\begin{gather}
	\im A_{3,2}^{R/L} = \frac{\im e^5}{(2\pi)^6} \int\frac{\diff^3k\,\diff^3k'}{\omega\, \omega'} 2 (p-k'-k)\cdot \varepsilon_q  \frac{\bar u_p^{R/L}   (\slashed{p}-\slashed{k}'-\slashed{k}) u_p^{R/L}}{\left[-2pk'-2pk+2kk'\right](2pk')(2pk)} 	-\frac{\im e^5}{(2\pi)^6} \int\frac{\diff^3k\,\diff^3k'}{\omega\, \omega'}  \frac{\bar u_p^{R/L}  \slashed\varepsilon_q  u_p^{R/L}}{(2pk')(2pk)} \punkt
\end{gather}	\end{widetext}
A small matrix calculation by hand or using Mathematica for the nominators gives
\begin{widetext}\begin{align}
	\im A_{3,2}^{R/L} = &\frac{\im e^5}{(2\pi)^6} \frac{4E}{\sqrt{1+\cos^2\theta_q}} \int\frac{\diff^3k\,\diff^3k'}{\omega\, \omega'}  \left[ \omega'\sin\theta'\cos\theta_q \e{\im\lambda_q(\phi'-\phi_q)} +  \omega\sin\theta\cos\theta_q \e{\im\lambda_q(\phi-\phi_q)} \nonumber \right. \\ 
	&\left. +E\sin\theta_q-\omega'\cos\theta'\sin\theta_q -\omega\cos\theta\sin\theta_q \right]  \frac{2E-\omega(1+\cos\theta)-\omega'(1+\cos\theta')}{\left[-2pk'-2pk+2kk'\right](2pk')(2pk)} \nonumber \\
	&-\frac{\im e^5}{(2\pi)^6} \frac{2E\sin\theta_q}{\sqrt{1+\cos^2\theta_q}} \int\frac{\diff^3k\,\diff^3k'}{\omega\, \omega'} \frac{1}{(2pk')(2pk)} \punkt \label{iA42phiphi}
\end{align}\end{widetext}
The $\int_0^{2\pi}\diff\phi \int_0^{2\pi}\diff\phi'$ integration of the first integral of \eqref{iA42phiphi} will make this term $0$. This can be seen by the following integrals, where as before the denominator is of the form $a+b\cos(\phi'-\phi)$:
\begin{gather} \int_0^{2\pi} \diff\phi\int_0^{2\pi}\diff\phi' \frac{\e{\im\lambda_q(\phi'-\phi_q)}}{a+b\cos(\phi'-\phi)} = 0 \komma \end{gather} 
\begin{gather} \int_0^{2\pi} \diff\phi\int_0^{2\pi}\diff\phi' \frac{\e{\im\lambda_q(\phi-\phi_q)}}{a+b\cos(\phi'-\phi)} = 0 \komma \end{gather}
\begin{gather} \int_0^{2\pi} \diff\phi\int_0^{2\pi}\diff\phi' \frac{1}{a+b\cos(\phi'-\phi)} = 0 \punkt \end{gather}
What is left over is the second line of \eqref{iA42phiphi} as result for the two-loop amplitude $\im A_{4}$:
\begin{widetext}\begin{gather}\label{iA42final}
	\im A_3^R = \im A_3^L = \im A_{3,2}^R=\im A_{3,2}^L = -\frac{\im e^5}{(2\pi)^6} \frac{2}{E}\frac{\sin\theta_q}{\sqrt{1+\cos^2\theta_q}} \log^2\left(\frac{E\delta}{m}\right)\left(\int\diff\omega\right)^2\komma
\end{gather}\end{widetext}
where we performed the angular integration of $\diff^3k\,\diff^3k'$ and we abbreviate $\int\diff\omega\int\diff\omega' = \left(\int\diff\omega\right)^2$. Thus, there is again no cancelation of the KLN anomaly in equation \eqref{ampsquare}.

This means that the two-loop amplitude $\im A_3$ from the diagrams in figure \ref{fig4} has only a $\log^2(m_e)$ divergence. Notice that this result \eqref{iA42final} is neither IR divergent, which is conform with \cite{Weinberg}, because in the forward limit the $B$-factor is $0$, nor there are single $\log(m_e)$ divergent terms.

{\bf Acknowledgements.}
We want to thank Gia Dvali and Sebastian Zell for many inspiring discussions and relevant and helpful comments. This work was supported  by the ERC Advanced Grant 339169 "Selfcompletion" and the grant FPA2015-65480-P. 

\end{document}